\pdfoutput=1
\documentclass[a4paper,oneside,11pt]{article}
\usepackage[margin=26mm]{geometry}

\usepackage[utf8]{inputenc}
\usepackage[T1]{fontenc}
\usepackage{fancyhdr}
\usepackage{lastpage}
\usepackage{titling}
\usepackage{caption}

\usepackage{authblk}
\usepackage[nohyperlinks,nolist]{acronym}
\usepackage{adjustbox}
\usepackage[ruled,linesnumbered,vlined]{algorithm2e}
\usepackage{amssymb}
\usepackage{amsmath}
\usepackage{booktabs}
\usepackage{csquotes}
\usepackage{enumitem}
\usepackage{listings}
\usepackage{tabularx}
\usepackage{xcolor}
\usepackage[
	backend=biber,
	safeinputenc,
	bibencoding=utf8,
	sorting=none,
	maxcitenames=10,
	maxbibnames=10,
	style=numeric,
	citestyle=numeric-comp]{biblatex}
\usepackage{hyperref}

\DontPrintSemicolon
\SetNlSty{texttt}{}{}
\SetAlgoNlRelativeSize{-1}
\SetNlSkip{1.2em}

\newcolumntype{Y}{>{\raggedleft\arraybackslash}X}

\newcommand{\ie}[1]{i.e.,}
\newcommand{\eg}[1]{e.g.,}

\newcommand{\noo}{-}
\newcommand{\yes}{$\checkmark$}

\newcommand{\pccoverage}{$t$-wise presence condition coverage}
\newcommand{\pcsampling}{$t$-wise presence condition sampling}
\newcommand{\magic}{\pccoverage{}}
\newcommand{\tartler}{Tartler et al.~\cite{TDS+:ATC14}}
\newcommand{\ruland}{Ruland et al.~\cite{RLB+:GPCE18}}
\newcommand{\prototype}{PRESICE}

\newcommand{\fm}{\mathcal{M}}
\newcommand{\mathfunction}[1]{\text{\textit{#1}}}

\newcommand{\refline}[1]{Line~\ref{#1}}

\begin{acronym}
\acro{spl}[SPL]{Software Product Line}
\end{acronym}

\usepackage{forest}
\usepackage{xcolor}
\usetikzlibrary{angles}
\definecolor{drawColor}{RGB}{128 128 128}
\newcommand{\circleSize}{0.25em}
\newcommand{\angleSize}{0.8em}

\forestset{
	/tikz/mandatory/.style={
		circle,fill=drawColor,
		draw=drawColor,
		inner sep=\circleSize
	},
	/tikz/optional/.style={
		circle,
		fill=white,
		draw=drawColor,
		inner sep=\circleSize
	},
	featureDiagram/.style={
		for tree={
			parent anchor = south,
			child anchor = north,
			draw = drawColor,
			edge = {draw=drawColor},
		}
	},
	/tikz/abstract/.style={
		fill = blue!85!cyan!5,
		draw = drawColor
	},
	/tikz/concrete/.style={
		fill = blue!85!cyan!20,
		draw = drawColor
	},
	mandatory/.style={
		edge label={node [mandatory] {} }
	},
	optional/.style={
		edge label={node [optional] {} }
	},
	or/.style={
		tikz+={
			\path (.parent) coordinate (A) -- (!u.children) coordinate (B) -- (!ul.parent) coordinate (C) pic[fill=drawColor, angle radius=\angleSize]{angle};
		}	
	},
	/tikz/or/.style={
	},
	alternative/.style={
		tikz+={
			\path (.parent) coordinate (A) -- (!u.children) coordinate (B) -- (!ul.parent) coordinate (C) pic[draw=drawColor, angle radius=\angleSize]{angle};
		}	
	},
	/tikz/alternative/.style={
	},
	/tikz/placeholder/.style={
	},
	collapsed/.style={
		rounded corners,
		no edge,
		for tree={
			fill opacity=0,
			draw opacity=0,
			l = 0em,
		}
	},
	/tikz/hiddenNodes/.style={
		midway,
		rounded corners,
		draw=drawColor,
		fill=white,
		minimum size = 1.2em,
		minimum width = 0.8em,
		scale=0.9
	},
}

\begin{document}

	\author[1]{Sebastian Krieter}
\author[1]{Thomas Thüm}
\author[2]{Sandro Schulze}
\author[3]{Sebastian Ruland}
\author[4]{Malte Lochau}
\author[2]{Gunter Saake}
\author[5]{Thomas Leich}
\affil[1]{University of Ulm, Ulm, Germany}
\affil[2]{University of Magdeburg, Magdeburg, Germany}
\affil[3]{TU Darmstadt, Darmstadt, Germany}
\affil[4]{University of Siegen, Siegen, Germany}
\affil[5]{Harz University of Applied Sciences, Wernigerode, Germany}

\title{T-Wise Presence Condition Coverage and Sampling for Configurable Systems}

\date{}
\maketitle
\thispagestyle{empty}

\begin{abstract}
\noindent
Sampling techniques, such as $t$-wise interaction sampling are used to enable efficient testing for configurable systems.
This is achieved by generating a small yet representative sample of configurations for a system, which circumvents testing the entire solution space.
However, by design, most recent approaches for $t$-wise interaction sampling only consider combinations of configuration options from a configurable system's variability model and do not take into account their mapping onto the solution space, thus potentially leaving critical implementation artifacts untested.
Tartler et al.\ address this problem by considering presence conditions of implementation artifacts rather than pure configuration options, but do not consider the possible interactions between these artifacts.
In this paper, we introduce \magic{}, which extends the approach of Tartler et al.\ by using presence conditions extracted from the code as basis to cover $t$-wise interactions.
This ensures that all $t$-wise interactions of implementation artifacts are included in the sample and that the chance of detecting combinations of faulty configuration options is increased.
We evaluate our approach in terms of testing efficiency and testing effectiveness by comparing the approach to existing $t$-wise interaction sampling techniques.
We show that \pcsampling{} is able to produce mostly smaller samples compared to $t$-wise interaction sampling, while guaranteeing a \pccoverage{} of 100\%.
\end{abstract}

	\section{Introduction}\label{SEC_Introduction}

Testing is an important task in software engineering to detect faults and to check intended behavior~\cite{ammann2016introduction,M10}.
However, exhaustive testing may be impossible and binds resources that could be used in other phases of development.
This is especially an issue when testing highly-configurable systems, such as \acp{spl}.
A product of an \ac{spl} can be derived from a configuration, which consist of a list of configuration options (\ie \emph{features}) that can be set either to \emph{true} or \emph{false}.
The entirety of all possible configurations for an \ac{spl} is called the \emph{configuration space}, which typically grows exponentially with the number of features~\cite{ER:IST11,LKL:SPLC12}.

A straight-forward testing strategy for \acp{spl} is product-based testing.
For this purpose, a set of products is derived from a set of different configurations and then test cases are run on each selected product respectively~\cite{TAK+:CSUR14}.
As testing every possible configuration in a product-based manner is usually not feasible due to an enormous configuration space, sampling strategies have been defined to generate a small yet representative set of configurations to test (\ie{} a configuration sample).
One such sampling strategy is \emph{$t$-wise interaction sampling}, which aims to generate a preferably small sample that covers all possible interactions of features of degree $t$~\cite{CDS:TSE08,MGSH:SPLC13}.
Using $t$-wise interaction sampling, developers can ensure that all interactions of at most $t$ features (\eg{} all selected, none selected, only one selected, etc.) are indeed contained in at least one configuration in the generated sample.
$T$-wise interaction sampling has shown to be a feasible trade-off between testing effectiveness and testing efficiency, as for small values of t (\ie{} $t \in \{2, 3\}$) it usually returns a relatively small sample, while also achieving reliable test results~\cite{AMS+:TOSEM18,KAR+:SPLC09,MGSH:SPLC13}.

A property of $t$-wise interaction sampling is that it works purely on the problem space of an \ac{spl}.
Thus, it is a black-box approach that does not take into account the mapping between features and actual implementation artifacts, such as source code, models, and test cases.
This can lead to a number of issues.
First, a traditional t-wise sampling algorithm may create samples that contain configurations that generate similar products, which can decrease overall testing efficiency.
Second, when choosing the parameter for $t$ the degree of faulty interactions is not known to the developers.
Thus, developers are incentivized to choose a low value for $t$ in order to keep the testing effort feasible.
However, some faults may require a feature interaction of a high degree to be included in  a configuration, and thus may not be included at all in a sample, which potentially decreases testing effectiveness.

Due to the black-box nature of $t$-wise interaction sampling, it may not reveal certain faults resulting from feature interaction beyond $t$ or requires too many configurations to even reach a certain code coverage or fault-detection rate.
To overcome this limitation, \tartler{} propose \emph{statement coverage}, a white-box approach that considers \emph{presence conditions} (\ie{} the selection of features for which an artifact is included in a product) to ensure that every artifact is present in at least one configuration in the sample and gets a chance of being tested.
However, simply including an implementation artifact in a product does not grantee that is indeed being tested.
Therefore, \ruland{} argue that, in addition to including each artifact, at least one test case must also cover each artifact in order to properly test it.
While the approach of \tartler{} can generate a relatively small sample, the approach of \ruland{} produces a relatively large sample, but guarantees that each artifact will indeed be tested.
Therefore, with this work, we aim to find a reasonable trade-off between both approaches by building on the approach of \tartler{} and try to increase testing effectiveness by combining it with t-wise interaction sampling.

We propose an extension of Tartler et al.{}'s coverage criterion \textit{$t$-wise presence condition coverage}, which combines $t$-wise interaction coverage with presence condition coverage of implementation artifacts.
Rather than counting only interactions between features, our new criterion considers interactions of presence conditions of implementation artifacts.
A sample with a $t$-wise presence condition coverage of 100\% ensures that every \emph{$t$-wise interactions of all implementation artifacts} is contained in at least one product and can thus be tested.
In addition, we present an algorithm for \textit{$t$-wise presence condition sampling}, which generates samples with a 100\% \textit{$t$-wise presence condition coverage} for a given configurable system and a given $t$.
This sampling algorithm works independently from the employed variability mechanisms such as preprocessors, build systems, feature-oriented programming, or aspect-oriented programming.
We implemented our algorithm within a prototype called \emph{\prototype{} (PRESence condItion CoveragE)} to investigate its testing effectiveness, testing efficiency, and sampling efficiency compared to traditional $t$-wise interaction sampling algorithms.
Our current prototype supports extraction of presence conditions from \acp{spl} that use the C preprocessor and the Kbuild build tool (\eg{} \texttt{BusyBox} and \texttt{Linux}).
Additionally, we extensively evaluate both, the coverage criterion and the algorithm using 27 real-world systems from different domains.
In summary, we contribute the following:

\begin{itemize}
	\item We propose a novel coverage criterion for product-based testing: $t$-wise presence condition coverage.
	\item We present a sampling algorithm for $t$-wise presence condition coverage.
	\item We provide an open-source implementation named named \prototype{} as part of FeatureIDE\footnote{\url{https://featureide.github.io/}}.
	\item We evaluate our coverage criterion and algorithm using 27 real-world systems.
	\item We publish all data from our experiments\footnote{\url{https://github.com/skrieter/evaluation-pc-sampling}}.
\end{itemize}

\section{Running Example and Problem Statement}

In the following, introduce a running example, which we use throughout the paper.
We use this example to describe the potential problems with the current approaches for $t$-wise interaction sampling and motivate our solution.

\subsection{Running Example}\label{sec_example}

\begin{figure}[t]
\begin{adjustbox}{max width=1.0\linewidth}
\sffamily
\begin{forest}
	featureDiagram
	[BUSYBOX\_TFTP,concrete[\textbf{T}FTP,concrete,optional][TFTPD,concrete,optional[TFTP\_\textbf{G}ET,concrete,optional][TFTP\_\textbf{P}UT,concrete,optional][TFTP\_\textbf{D}EBUG,concrete,optional][TFTP\_\textbf{B}LOCKSIZE,concrete,optional]]]	
	\matrix [anchor=north west] at (current bounding box.north east) {
		\node [placeholder] {}; \\
	};
	\matrix [draw=drawColor,anchor=north west] at (2.65,0.32) {
		\node [label=center:\underline{Legend:}] {}; \\
		\node [optional,label=right:Optional] {}; \\
	};
\end{forest}
\end{adjustbox}
\caption{Excerpt of \emph{BusyBox} feature diagram.}
\label{fig_busbox_fm}
\end{figure}
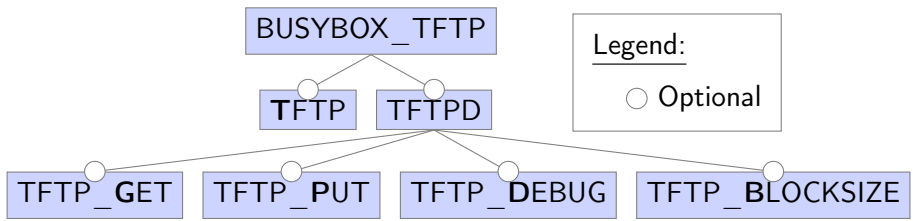

To illustrate the challenge of finding effective and efficient samples, we show a slightly edited code snippet from the system \texttt{BusyBox} in \autoref{lst_ifexample}, which uses the C preprocessor~\cite{stallman1987c} to implement its variability~\cite{LAL+:ICSE10}.
The example is taken from the file \texttt{tftp.c}, which handles client-server communication via tftp.
The code contains five features, \texttt{TFTP} $(T)$, \texttt{TFTP\_GET} $(G)$, \texttt{TFTP\_PUT} $(P)$, \texttt{TFTP\_BLOCKSIZE} $(B)$, and \texttt{TFTP\_DEBUG} $(D)$, which each can be set to either \textit{true} or \textit{false}.
We display their dependencies in \autoref{fig_busbox_fm} using an excerpt of the \texttt{BusyBox} variability model, represented as a feature diagram.
The additional features \texttt{BusyBox\_TFTP} $(BB)$ and \texttt{TFTPD} $(TD)$ do not appear in the code snippet, but are necessary to visualize the feature diagram hierarchy.
In the remainder of the paper, we use the provided abbreviations of the feature names to ease the readability of all propositional formulas using these features.

In the example, we changed the statement in \autoref{lst_line_error} such that a compilation error occurs for certain products.
The variable \texttt{blksize} is declared in \autoref{lst_line_decl}, which is dependent on feature $B$.
Then, \texttt{blksize} is used in \autoref{lst_line_error}, which is dependent on feature $D$.
Thus, if $D$ is selected in a configuration, but $B$ is not, the generated product will be syntactically incorrect.

\begin{figure}
{\renewcommand\thelstnumber{%
\ifcase\value{lstnumber} 0
\or 1%
\or %
\or 621%
\or 622%
\or %
\or 625%
\or 626%
\or %
\or 649%
\or 650%
\or 651%
\or %
\or 670%
\or 671%
\or 672%
\or %
\or 690%
\or 691%
\or %
\or 827%
\fi
}
\begin{lstlisting}[
belowcaptionskip=0pt,
belowskip=0pt,
numbersep=5pt
]
#if TFTP_GET || TFTP_PUT // @$G \vee P$@ @\label{lst_line_or}@
// ...
# if TFTP // @$T$@
int tftp_main(int argc, char **argv) {@\label{lst_line_function}@
	// ...
	#  if TFTP_BLOCKSIZE // @$B$@
	const char *blksize_str=TFTP_BLKSIZE_DEFAULT; @\label{lst_line_decl}@
	// ...
	int blksize = tftp_blksize_check(blksize_str, 65564);
	if (blksize < 0) return EXIT_FAILURE;
	#  endif
	// ...
	#  if TFTP_DEBUG // @$D$@
	printf("blksize = %d\n", blksize); // changed @\label{lst_line_error}@
	#  endif
	// ...
}
# endif
// ...
#endif
	\end{lstlisting}}
	\caption{Example adopted from \texttt{BusyBox}.}
	\label{lst_ifexample}
\end{figure}

\subsection{Problem Statement}\label{sec_problem}

$T$-wise interaction coverage only considers interactions between single features, and thus is based purely within the problem space of an \ac{spl}.
This can lead to two flaws.
First, $t$-wise interaction coverage may consider some irrelevant feature interactions, which can yield a large sample, potentially leading to a low testing efficiency.
Second,  $t$-wise interaction coverage may not be sufficient to reliably detect a fault resulting from an interaction of a degree larger than $t$, potentially leading to a low testing effectiveness.
In order to reliably find faults with a high interaction degree a high value for $t$ is required.
However, traditional $t$-wise interaction coverage does not scale well with higher values for t, as the number of possible feature interactions grows exponentially with the parameter $t$.
Thus, often a low value, such as $t = 2$ or $t = 3$, is chosen, which keeps the testing effort manageable, but also decreases the fault detection rate.
In the following, we explain both flaws in more detail.
With the introduction of our new coverage criterion, we aim to address both of these flaws by taking the solution space into account.

\paragraph{Efficiency of Sample-Based Testing}
In sample-based testing, we run the all test cases of a system once per sampled configuration.
Although the execution time of the test cases may differ from configuration to configuration, in general, if the number of configurations in a sample (\ie{} the sample size) increases the overall testing effort increases as well.
Thus, analogous to other research, we consider the testing of smaller samples to be more efficient (\ie{} \emph{testing efficiency}~\cite{RLB+:GPCE18}).

Applying pair-wise interaction sampling to our running example considers every interaction between two features.
For instance, in \autoref{lst_ifexample}, for the features $P$ and $G$, all four possible interactions are considered.
However, the three interactions $(P, \neg G)$, $(\neg P, G)$, $(P, G)$ all lead to the same product, as all of them satisfy the expression in \autoref{lst_line_or}.
Thus, it is sufficient to consider only two interactions (\eg{} $(\neg P, \neg G)$ and $(P, G)$) for this code snippet.
In the table below, we show the sample generated from the $t$-wise interaction sampling algorithm ICPL~\cite{JHF:MODELS11,MHF:SPLC12}, which consists of six configurations:
\vspace{4pt}
\begin{center}
	\small
	\begin{tabular}{l|cccccc}
		\toprule
		\textbf{Feature} & \multicolumn{6}{c}{\textbf{Configurations}} \\ 
		& \textbf{01} & \textbf{02} & \textbf{03} & \textbf{04} & \textbf{05} & \textbf{06} \\ 
		\midrule
		\texttt{TFTP (T)} 				& \noo & \yes & \noo & \yes & \yes & \noo \\ 
		\texttt{TFTP\_GET (G)}			& \noo & \yes & \yes & \noo & \yes & \noo \\ 
		\texttt{TFTP\_PUT (P)} 			& \noo & \yes & \noo & \noo & \yes & \yes \\ 
		\texttt{TFTP\_DEBUG (D)}		& \noo & \noo & \yes & \noo & \yes & \yes \\ 
		\texttt{TFTP\_BLOCKSIZE (B)}	& \noo & \yes & \yes & \noo & \noo & \yes \\ 
		\bottomrule 
	\end{tabular}
\end{center}
\vspace{4pt}
\noindent
All four interactions of $P$ and $G$ are included.
Including such unnecessary interaction can lead to a larger sample, and thus to a lower testing efficiency.

\paragraph{Effectiveness of Sample-Based Testing}\label{subsec_example_effectiveness}
We consider a sample to be more effective, the more faults could be detected using its derived products.
Analogous to other research, we refer to this as \emph{testing effectiveness}~\cite{RLB+:GPCE18}.

Although the fault in \autoref{lst_ifexample} apparently involves only two features, it is in fact a feature interaction of degree four.
To actually generate a product that contains the error, the corresponding configuration must have the feature $B$ deselected and the features $T$, $D$, and $G$ or $P$ selected.
Below, we show the sample generated by the pair-wise sampling algorithm IncLing~\cite{AKT+:GPCE16}:
\vspace{4pt}
\begin{center}
	\small
	\begin{tabular}{l|ccccccc}
		\toprule
		\textbf{Feature} & \multicolumn{7}{c}{\textbf{Configurations}} \\ 
		& \textbf{01} & \textbf{02} & \textbf{03} & \textbf{04} & \textbf{05} & \textbf{06} & \textbf{07} \\ 
		\midrule
		\texttt{TFTP (T)} 				& \yes & \noo & \yes & \noo & \noo & \noo & \yes \\ 
		\texttt{TFTP\_GET (G)} 			& \yes & \noo & \noo & \yes & \noo & \noo & \yes \\ 
		\texttt{TFTP\_PUT (P)} 			& \yes & \noo & \noo & \yes & \noo & \yes & \noo \\ 
		\texttt{TFTP\_DEBUG (D)} 		& \yes & \noo & \noo & \noo & \yes & \yes & \noo \\ 
		\texttt{TFTP\_BLOCKSIZE (B)}	& \yes & \noo & \yes & \noo & \yes & \noo & \noo \\ 
		\bottomrule 
	\end{tabular}
\end{center}
\vspace{4pt}
\noindent
Although the interaction $(D, \neg B)$ is covered in Configuration 06, the actual fault will not be included in the product, as the feature $T$ is not selected and the preprocessor will remove the entire code block.
Thus, the testing effectiveness of this sample is decreased.

\section{Foundations of Configurable Systems}\label{sec_background_fm}

Before we can properly introduce the concept of \pccoverage{}, we have to provide some essential definitions on feature modeling, configurations, and presence conditions.
Our approach takes a feature model and a list of presence conditions as input and generates a list of configurations (\ie{} a sample).
Therefore, in the following, we revisit the basic notion of feature models, configurations, and presence conditions.
We introduce all notions using propositional formulas and set notation.

\subsection{Feature Models}
A feature model specifies all features of an \ac{spl} and their interdependencies~\cite{CN01,PBL05,ABKS13}.
We define a feature model $\fm = (\mathcal{F}, \mathcal{D})$ as a tuple, consisting of a set of features $\mathcal{F}$ and a set of dependencies $\mathcal{D}$ on $\mathcal{F}$.
All features of a feature model are contained in the set $\mathcal{F} = \{F_1, ..., F_n\}$, where $n$ is the total number of features.
We represent the dependencies of a feature model as clauses of a propositional formula in conjunctive normal form (CNF).
Each dependency in $\mathcal{D}$ represents one such clause over $\mathcal{F}$ (\ie{} $\mathcal{D} = \{D_1, ..., D_m\}$, where $m$ is the number of clauses).
We denote a clause as a set of literals over $\mathcal{F}$.
A literal is either a feature from $\mathcal{F}$ (\ie{} a positive literal) or a negated feature from $\mathcal{F}$ (\ie{} a negative literal).
We define the function $\mathcal{L}$ that provides the set of literals for a feature model, $\mathcal{L}(\fm) = \{false, true, \neg F_{1}, ..., \neg F_{n}, F_{1}, ..., F_{n}\}$.
For example, consider the features \texttt{TFTP\_PUT} $(P)$, \texttt{TFTPD} $(TD)$, and \texttt{BUSYBOX\_TFTP} $(BB)$ from \autoref{fig_busbox_fm}.
Their dependencies can be written as the two clauses $D_1 = \neg P \vee TD$ and $D_2 = \neg TD \vee BB$.

\subsection{Configurations}\label{sec_background_configuration}

A configuration represents a selection of features from a feature model.
From a configuration, we can derive the corresponding product using the variability mechanism of the \ac{spl}~\cite{CN01,PBL05,ABKS13}.
We define a configuration as a set of literals $C$, such that $C \subseteq \mathcal{L}(\fm)$ with $\forall l \in \mathcal{L}(\fm) : l \notin C \vee \neg l \notin C$.
If a literal is contained in a configuration, the corresponding feature is defined as either selected (positive literal) or deselected (negative literal).

If all features are defined within a configuration, we call it \textit{complete} and otherwise \textit{partial}.
We define this with the function:
\begin{equation*}
    \mathfunction{complete}(C, \fm) =
	\begin{cases}
		\mathfunction{true}	& |C| = |\mathcal{F}| \\
		\mathfunction{false}	& \text{otherwise} \\
	\end{cases}
\end{equation*}

A clause of a feature model represents a disjunction of literals.
Thus, if a configuration contains at least one literal from a clause in $\mathcal{D}$, it satisfies this clause.
Contrary, if a configuration contains all complementary literals of a clause, it contradicts this clause and, hence the entire feature model.
Thus, if a configuration contradicts at least one clause, we call it \textit{invalid}.
We call a configuration \emph{valid}, if it allows all clauses of a feature model to be satisfied:
\begin{equation*}
    \mathfunction{valid}(C, \fm) =
	\begin{cases}
		\mathfunction{true}	& \exists C' \supseteq C : \forall D \in \mathcal{D} : C' \cap D \neq \emptyset \\
		\mathfunction{false}	& \text{otherwise} \\
	\end{cases}
\end{equation*}
Note that, a partial configuration may also neither satisfy nor contradict a clause.
In particular, a valid configuration may be partial and is not required to satisfy all clauses, as long as all clauses \emph{can be satisfied} by adding more literals to the configuration.

\subsection{Presence Conditions}\label{sec_background_pcs}

A \emph{presence condition} is a propositional formula over a feature model that describes whether a certain aspect is true (\ie{} present) for a given configuration and/or its respective product.
A presence condition is \emph{not} part of a feature model and does \emph{not} limit the valid configuration space.
Rather it describes a subset of the valid configuration space (\ie{} the set of all valid configurations that satisfy its propositional formula) for which a certain aspect is true.
In this paper, we use presence conditions to describe whether an implementation artifacts is present in a product or not.
In particular, we are interested in single lines of code (\ie{} statements).
Furthermore, in our evaluation, we also use presence conditions to describe whether a particular fault is present in a product or not.
If and only if a configuration contains a combination of feature selections that satisfy a presence condition, the corresponding implementation artifact is included in the respective product of that configuration.
If a presence condition is satisfied by a configuration we call it \emph{active} and otherwise \emph{inactive}.

For instance, \autoref{lst_line_function} of \autoref{lst_ifexample} has the presence condition $(G \vee P) \wedge T$, which can be derived from the nested C preprocessor annotations within the source code.
This means that \autoref{lst_line_function} is present in a product if the corresponding configuration has the feature $G$ or $P$ selected and the feature $T$ selected, because the presence condition is active for these configurations.
When we use presence conditions, in most cases, we are interested in which feature selections would make a presence condition active.
In the example above these are the combinations $G \wedge T$ and $P \wedge T$, if any of these two is true, the presence condition is active.
For this reason, it is beneficial to represent presence conditions in \emph{disjunctive normal form} (DNF), which consists of a disjunction of conjunctive clauses.
In a DNF, each clause represents a feature selection that satisfies the presence condition.
In order to include an artifact in a product, at least one clause of the DNF must be satisfied, which means that all of its literals must be contained within a configuration.
For example, we can write the presence condition of \autoref{lst_line_function} as $\mathcal{P}^{+}_{\ref*{lst_line_function}} = (G \wedge T) \vee (P \wedge T)$.
Thus, in order to include this line in a product, the corresponding configuration must either contain features $G$ and $T$ or $P$ and $T$.

To represent presence conditions within our formalism, we use a similar notation as for dependencies.
We define a presence condition such that $\mathcal{P} = \{ P_1, ..., P_k \}$ with $P_i \subseteq \mathcal{L}(\fm)$.
In the context of \pccoverage, we refer to a DNF clause simply as clause.
Further, we define the function \textit{active} to check whether a presence condition is satisfied for a valid configuration:
\begin{equation*}
    \mathfunction{active}(\mathcal{P}, C) =
	\begin{cases}
		\mathfunction{true}	& \exists P \in \mathcal{P} : P \subseteq C \\
		\mathfunction{false}	& \text{otherwise} \\
	\end{cases}
\end{equation*}

	\section{Presence Condition Coverage}

\newcommand{\result}{\mathcal{C}_{Sample}}
\newcommand{\interaction}{\mathcal{P}_I}

In the following, we define our coverage criterion for \pccoverage{}, based on presence conditions, the feature model, the parameter $t$, and a configuration sample.
To this end, we describe interactions between presence conditions and how our coverage criterion is calculated for a given sample and $t$.

\subsection{Presence Condition Interactions}\label{sec_concept_interactions}

An interaction between two presence conditions (\ie{} between their implementation artifacts) is similar to interactions between features.
However, the key difference is that a presence condition can be an arbitrary propositional formula.
In a complete configuration, a feature can either be selected, if its corresponding positive literal is included in the configuration, or deselected, if its negative literal is included.
In contrast, a presence condition may be active or inactive for several different literal combinations (cf. \autoref{sec_background_pcs}).
Thus, an interaction between $t$ presence conditions must include all possible union sets of these literal combinations.

\paragraph{Inactive Presence Conditions}
In order to account for the interactions between \emph{present} and \emph{absent} implementation artifacts, we must consider the interactions between \emph{active} and \emph{inactive} presence conditions.
For this reason, we construct the complement of each presence condition, which is itself a presence condition and represents all literal combinations, for which the original presence condition is inactive.
To this end, we negate the formula of a presence condition and convert it back into a DNF.
As an example, consider the presence condition $\mathcal{P}^{+}_{\ref*{lst_line_decl}} = (G \wedge T \wedge B) \vee (P \wedge T \wedge B)$ for \autoref{lst_line_decl} of \autoref{lst_ifexample}.
We process this formula as follows:
\begin{alignat*}{3}
	\mathcal{P}^{-}_{\ref*{lst_line_decl}} &=\;&&\neg\,\mathcal{P}^{+}_{\ref*{lst_line_decl}}\\
	&=\;&&\neg ( (G \wedge T \wedge B) \vee (P \wedge T \wedge B) )\\
	_{De~Morgan's~Law}&\equiv~&& (\neg G \vee \neg T \vee \neg B) \wedge (\neg P \vee \neg T \vee \neg B)\\
	_{Distributive~Law}&\equiv~&& (\neg G \wedge\neg P) \vee (\neg T) \vee (\neg B)
\end{alignat*}
After negating the formula, we apply De Morgan's law to get a CNF and then apply the distributive law to convert it back into a DNF.
While the application of the distributive law could theoretically lead to an exponential growth in the number of clauses, in practice a presence condition consists of only a few clauses and literals, which makes the computational effort for this transformation neglectable.
From the new DNF $\mathcal{P}^{-}_{\ref*{lst_line_decl}}$, we see that if a configuration either contains the literal $\neg T$, $\neg B$, or both, $\neg G$ and $\neg P$, the resulting product will not include \autoref{lst_line_decl}.

For a given \ac{spl}, we construct one set $\mathcal{P\!C}$ that contains DNFs of all presence conditions from the \ac{spl} and their corresponding complements.
To construct all $t$-wise interactions for a given \ac{spl}, we can then generate all $t$-wise combinations of $\mathcal{P\!C}$ by constructing the Cartesian product for the given $t$ (\ie{} $\mathcal{P\!C}^{t}$).

\paragraph{Combined Presence Condition}
For each interaction of $t$ presence conditions, we can build a new combined presence condition $\mathcal{P}_I$ that is satisfied if and only if all individual presence conditions are active or inactive, respectively.
To this end, we conjoin the DNFs of all presence conditions involved in a given interaction and converting the resulting expression back to a DNF.
For each presence condition that is active in the given interaction we use its original DNF and for each presence condition that is inactive we use its complementary DNF.
For instance, consider the DNFs for \emph{excluding} \autoref{lst_line_decl} and \emph{including} \autoref{lst_line_error} of \autoref{lst_ifexample}:
\begin{alignat*}{2}
	&\text{\autoref{lst_line_decl}:~~~}&&\mathcal{P}^{-}_{\ref*{lst_line_decl}} = (\neg G \wedge\neg P) \vee (\neg T) \vee (\neg B)\\
	&\text{\autoref{lst_line_error}:}&&\mathcal{P}^{+}_{\ref*{lst_line_error}} = (G \wedge T \wedge D) \vee (P \wedge T \wedge D)
\end{alignat*}
We build the presence condition for the interaction ($\mathcal{P}_{I}$) by conjoining the literals of each pair-wise clause combination:
\begin{alignat*}{1}
	\mathcal{P}_{I} =\;&\mathcal{P}^{-}_{\ref*{lst_line_decl}} \wedge \mathcal{P}^{+}_{\ref*{lst_line_error}}\\
	_{Distributive~Law}\equiv\;&	((\neg G \wedge\neg P) \wedge (G \wedge T \wedge D)) \vee 
	((\neg G \wedge\neg P) \wedge (P \wedge T \wedge D)) \vee\\
	&((\neg T)  \wedge (G \wedge T \wedge D)) \vee 
	((\neg T)  \wedge (P \wedge T \wedge D)) \\
	&((\neg B)  \wedge (G \wedge T \wedge D)) \vee 
	((\neg B)  \wedge (P \wedge T \wedge D)) \\
	_{simplify}\equiv\;&(\neg B \wedge G \wedge T \wedge D) \vee (\neg B \wedge P \wedge T \wedge D)
\end{alignat*}
After merging, we further simplify the combined DNF by removing contradictions and redundant clauses.
In case of our example, this results in the simplified DNF $(\neg B \wedge G \wedge T \wedge D) \vee (\neg B \wedge P \wedge T \wedge D)$.
Thus, if a configuration either contains the literals $\neg B$, $G$, $T$, and $D$ or $\neg B$, $G$, $T$, and $P$ \autoref{lst_line_error} will appear in the resulting product, but not \autoref{lst_line_decl}.

\subsection{Presence Condition Coverage Criterion}

Given a set of presence conditions ($\mathcal{P\!C}$), feature model ($\fm$), $t$, and configuration sample ($\mathcal{C}$) for an \ac{spl}, we can determine a value for our coverage criterion \emph{\pccoverage{}}.
We define \pccoverage{} as the ratio between the number of $t$-wise presence condition interactions that are covered by $\mathcal{C}$ and the number of valid $t$-wise presence condition interactions for $\fm$.
Naturally, with \pcsampling{} we aim to compute a sample that achieves a coverage of 100\%.

In our running example, there are five different presence conditions and their five complements. 
\begin{alignat*}{2}
&true, && false,\\
&(G) \vee (P), && (\neg G \wedge \neg P),\\
&(G \wedge T) \vee (P \wedge T), && (\neg G \wedge \neg P) \vee (\neg T),\\
&(G \wedge T \wedge B) \vee (P \wedge T \wedge B),~ && (\neg G \wedge \neg P) \vee (\neg T) \vee (\neg B),\\
&(G \wedge T \wedge D) \vee (P \wedge T \wedge D),~ && (\neg G \wedge \neg P) \vee (\neg T) \vee (\neg D)
\end{alignat*}
In total, there are 11 valid combinations.
Considering the sample from IncLing in \autoref{subsec_example_effectiveness}, there are two interactions that are not covered by any configuration in the sample: $(G \wedge T \wedge B \wedge \neg D) \vee (P \wedge T \wedge B \wedge \neg D)$ and $(G \wedge T \wedge \neg B \wedge D) \vee (P \wedge T \wedge \neg B \wedge D)$.
Thus, the number of interactions covered by the sample is 9, which results in a pair-wise presence condition coverage of approximately $82\%$.

\section{T-Wise Presence Condition Sampling}\label{sec_concept}

In the following, we describe our algorithm to achieve \pccoverage{}.
The algorithm consists of three steps.
First, we extract the presence conditions for each line of code from the given \ac{spl}.
Second, we construct the set $\mathcal{P\!C}$ by \emph{preprocessing the presence conditions} extracted from the \ac{spl}, removing tautologies, contradictions, and equivalent conditions and computing the complementary presence conditions (cf. \autoref{sec_concept_interactions}).
Third, we iteratively \emph{construct a configuration sample} that includes all valid $t$-wise combinations of $\mathcal{P\!C}$.

\subsection{Extracting Presence Conditions}\label{sec_epp}

As a first step, we have to extract a list of presence conditions from the target \ac{spl}.
This step is highly dependent on the specifics of the given \ac{spl}, such as the programming language, the variability mechanism, and the configuration mechanism.
Therefore, a general approach for this step is out of scope for this paper.
In fact, both processes, extracting presence conditions and testing products, are independent from our approach and can be adapted to any suitable variability mechanism, programming language, and testing framework.
We did however implement an algorithm for extracting presence conditions from \acp{spl} that use the language C and the C preprocessor as a variability mechanism.
The algorithm is based on the tool PCLocator~\cite{KuiterKKLLS18}, which is able to parse C files and analyze preprocessor annotations.
The basic procedure from this algorithm would also work for other preprocessor implementations, such as Antenna for Java, but would of course require a different parser implementation.
In the following, we describe the general logic behind the algorithm.
We give some more details for our particular implementation in our prototype \prototype{} in \autoref{sec_implementation}.

In general, the algorithm parses a C file and determines a presence condition for each line.
As a result the algorithm returns a list of presence conditions for each file.
The C preprocessor annotations that are used for variability management always form a block around variable code artifacts.
Each block begins with an annotation, such as \texttt{\#if} or \texttt{\#ifdef} and ends with an annotation, such as \texttt{\#else} or \texttt{\#endif}.
These block can be nested, as we display in our example in \autoref{lst_ifexample}.
Consequently, for each line in a C file, the algorithm determines the C preprocessor blocks that surrounds the line.
The presence condition of the line is then constructed by the conjunction of the conditionals of the surrounding blocks.

Due to the high expressiveness of the C preprocessor, the extraction algorithm has some limitations, which can lead to incorrect results in some rare cases.
Most notably, the algorithm does not expand any preprocessor macro statements, which can lead parsing problems or missed annotations.
In case of a parsing problem for a line, the impacted presence conditions are assumed to be true (\ie{} the lines are always present).
Further, the algorithm does only consider Boolean features.
All features with numerical or string values are ignored.
Due to these limitations, it is possible that some returned presence condition might be incorrect or incomplete.

\subsection{Preprocessing Presence Conditions}\label{sec_ppp}

Our core algorithm for \pcsampling{} expects as input a list of presence conditions $\mathcal{P\!C}$, a feature model $\fm$, and a value for $t$.
From the previous step, we get a raw list of extracted presence conditions from an \ac{spl}, which can be arbitrary propositional formulas in any order.
Therefore, the next step is to preprocess this initial list.
In detail, we convert all presence conditions into DNF (cf. \autoref{sec_background_pcs}), compute their complements (cf. \autoref{sec_concept_interactions}), and clean up the list by removing equivalent formulas, tautologies, and contradictions.
In the end, we converted all raw presence conditions into a minimal list of DNFs $\mathcal{P\!C}$.
Note that we do not need to convert a presence condition from CNF to DNF, but directly convert the propositional formulas extracted from the implementation artifacts, which typically contain only a small set of features and no complicated formulas.

Regarding our example in \autoref{lst_ifexample}, we extract the following raw presence conditions (for brevity, we already omit duplicates):
\begin{align*}
&true,~~G \vee P,~~(G \vee P) \wedge T,~~(G \vee P) \wedge T \wedge B,~~(G \vee P) \wedge T \wedge D
\end{align*}
The preprocessing results in the following list:
\begin{align*}
&(G) \vee (P),~~(G \wedge T) \vee (P \wedge T),~~(G \wedge T \wedge B) \vee (P \wedge T \wedge B),\\
&(G \wedge T \wedge D) \vee (P \wedge T \wedge D), (\neg G \wedge \neg P),~~(\neg G \wedge \neg P) \vee (\neg T), \\
&(\neg G \wedge \neg P) \vee (\neg T) \vee (\neg B),~~(\neg G \wedge \neg P) \vee (\neg T) \vee (\neg D)
\end{align*}

Splitting our algorithm in a preprocessing and a sampling part allows us to flexibly determine the input for the actual sampling algorithm.
For instance, we can use the sampling algorithm also for regular $t$-wise interaction sampling, if we construct $\mathcal{P\!C}$ accordingly (\ie{} $\mathcal{P\!C} = \{L \subseteq \mathcal{L}(\fm)~|~|L| = 1$).
Furthermore, it also allows us to build subsets of $\mathcal{P\!C}$ and only consider interactions between these.
To this end, we run the sampling algorithm once for each subset and iteratively add configurations to $\result$.
Using this, we can, for example, group presence conditions for single files or folders, and thus limit the number of possible interactions and the resulting sample size.

\subsection{Constructing a Configuration Sample}

After computing the set $\mathcal{P\!C}$, we continue with the third step of our algorithm, constructing a sample of valid and complete configurations $\result$.
For this, we extend the existing algorithm YASA~\cite{KTSSL20}, a deterministic greedy algorithm.
We start with an empty sample $\result$ and then iterate over all $t$-wise interactions one at a time.
For each, we either add a new partial configuration to the sample or the literals of one clause of the interaction's presence condition to an existing configuration.
Due to the nature of this algorithm, we are guaranteed to check every possible combination of the given presence conditions exactly once.
For each combination or presence conditions, we either include it in at least one configuration in the sample or determine that it cannot be covered by any valid configuration.
Since there are finitely many possible combinations (\ie{} $2^t \cdot \binom{k}{t}$ with $k$ being the number of presence conditions), the algorithm always terminates and guarantees that a presence condition coverage of 100\% is achieved by the computed sample.

We present pseudocode of the second step in \autoref{algo:main}.
It takes a parameter $t$, the feature model $\fm$, and a list of DNFs $\mathcal{P\!C}$ and returns a sample $\result$ that covers every $t$-wise combination of DNFs in $\mathcal{P\!C}$.
At the start, we initialize $\result$ with an empty set (\refline{line_init}).
To list all presence condition interactions up to degree $t$, we then build the Cartesian product for $\mathcal{P\!C}$ with itself $t$ times (\ie{} $\mathcal{P\!C}^{t}$) (\refline{line_combination_loop}).
For each interaction in $\mathcal{P\!C}^{t}$, we compute the combined presence condition $\mathcal{P}_{I}$ as outlined in \autoref{sec_concept_interactions} (\refline{line_merge}).
Using $\mathcal{P}_{I}$, we check whether this interaction is already covered by at least one configuration in the sample (\refline{line_not_covered}).
If not, we try to cover the interaction by iterating over all clauses in $\mathcal{P}_{I}$ (\refline{line_clause_loop}).
For each clause $P$, we first check, whether it is satisfiable with regard to the feature model $\fm$  (\refline{line_clause_validity}).
If so, we add it to a temporary set of valid clauses $\mathcal{P}_{valid}$  (\refline{line_add_p}).
Second, we iterate over all configurations in our current sample and check, whether we can add the literals of $P$ to it without causing a contradiction (\refline{line_config_validity}).
If it does not cause a contradiction, we add all literals to the configuration (\ie{} covering the interaction) and continue with the next interaction (\refline{line_add_combination} and \ref{line_continue}).
Otherwise, if we cannot find any suitable configuration for any clause in $\mathcal{P}_{I}$, we use the smallest clause in $\mathcal{P}_{valid}$, if any, and use it to build a new configuration that we add to our sample (\refline{line_add_config1} and \ref{line_add_config2}).

\IncMargin{1em}
\begin{algorithm}
	\KwData{Presence Conditions $\mathcal{P\!C}$, Feature Model $\fm$, Parameter $t$}
	\KwResult{List of configurations $\result$}
	$\result \leftarrow \emptyset$\; \label{line_init}
	\For{$I \in \mathcal{P\!C}^{t}$}{ \label{line_combination_loop}
		$\mathcal{P}_{I} \leftarrow \{\bigcup_{P \in T} P~|~T \in \prod_{\mathcal{P} \in I} \mathcal{P}\}$\; \label{line_merge}
		\If{$\nexists C \in \result: \mathfunction{active}(\mathcal{P}_{I}, C)$} { \label{line_not_covered}
			$\mathcal{P}_{valid} \leftarrow \emptyset$\;
			\For{$P \in \mathcal{P}_{I}$}{ \label{line_clause_loop}
				\If{$\mathfunction{valid}(P, \fm)$} { \label{line_clause_validity}
					\For{$C \in \result$}{ \label{line_config_loop}
						\If {$\mathfunction{valid}(C \cup P, \fm)$}{ \label{line_config_validity}
							$C \leftarrow C \cup P$\; \label{line_add_combination}
							\textbf{continue} \autoref*{line_combination_loop}\; \label{line_continue}
						}
					}
					$\mathcal{P}_{valid} \leftarrow \mathcal{P}_{valid} \cup P$\; \label{line_add_p}
				}
			}
			\If{$\mathcal{P}_{valid} \neq \emptyset$} { \label{line_add_config1}
				$\result \leftarrow \result \cup min(\mathcal{P}_{valid})$\;  \label{line_add_config2}
			}
		}
	}
	\KwRet $\result$\;
	\caption{Pseudocode for Constructing a Sample}
	\label{algo:main}
\end{algorithm}%
\DecMargin{1em}

As an example, we describe some iterations of the algorithm for \autoref{lst_ifexample}.
In the first iteration we get the combination $I = ((G \vee P), (\neg G \wedge \neg P))$ (\refline{line_combination_loop}), which is converted into the combined presence condition $\mathcal{P}_{I} = G \wedge \neg G \vee P \wedge \neg P$ (\refline{line_merge}).
As there is no configuration yet in $\result$, we continue (\refline{line_not_covered}).
Both clauses in $\mathcal{P}_{I}$, $P_1 = G \wedge \neg G$ and $P_2 = P \wedge \neg P$ are invalid and are therefore not considered for inclusion into a configuration (\refline{line_clause_validity}).
Thus, we continue with the next iteration.
In the second iteration we get the combination $I = ((G \vee P), ((G \wedge T) \vee (P \wedge T)))$ (\refline{line_combination_loop}), which results in $\mathcal{P}_{I} = (G \wedge T) \vee (P \wedge T)$ (\refline{line_merge}).
There is still no configuration in $\result$, which may cover this interaction (\refline{line_not_covered}).
Both clauses $P_1 = G \wedge T$ and $P_2 = P \wedge T$ are valid  (\refline{line_clause_validity}), but there is no configuration yet to which they could be added (\refline{line_config_loop}).
Thus, both of them are added to $\mathcal{P}_{valid}$ (\refline{line_add_p}).
Next, the smallest clause in $\mathcal{P}_{valid}$ is added to $\result$ (\refline{line_add_config2}).
As both clauses have the same size, we use the first one, resulting in  $\result = \{\{G, T\}\}$.
In the third iteration we get combined presence condition $\mathcal{P}_{I} = (G \wedge T \wedge D) \vee (P \wedge T \wedge D)$.
Both clauses are valid and can be added to the existing configuration in $\result$ (\refline{line_config_validity}).
We use the first clause, which results in $\result = \{\{G, T, D\}\}$ (\refline{line_add_combination}).
In the fourth iteration we get combined presence condition $\mathcal{P}_{I} = (G \wedge T \wedge \neg D) \vee (P \wedge T \wedge \neg D)$.
Here, both clauses are valid, but conflict with the existing configuration in $\result$ (\refline{line_config_validity}).
Thus, a new configuration is added, It can be added to the existing configuration (\refline{line_add_config2}), resulting in $\result = \{\{G, T, D\}, \{G, T, \neg D\}\}$.

The complete configuration sample after the algorithm terminated is shown in the following table:
\vspace{4pt}
\begin{center}
	\small
	\begin{tabular}{l|ccccc}
		\toprule
		\textbf{Feature} & \multicolumn{5}{c}{\textbf{Configurations}} \\ 
		& \textbf{01} & \textbf{02} & \textbf{03} & \textbf{04} & \textbf{05} \\ 
		\midrule
		\texttt{TFTP (T)}				& \yes & \yes & \yes & \noo & \noo \\ 
		\texttt{TFTP\_GET (G)}			& \yes & \yes & \yes & \yes & \noo \\ 
		\texttt{TFTP\_PUT (P)}			& \yes & \yes & \yes & \yes & \noo \\ 
		\texttt{TFTP\_DEBUG (D)}		& \yes & \noo & \yes & \noo & \yes \\ 
		\texttt{TFTP\_BLOCKSIZE (B)}	& \yes & \yes & \noo & \noo & \yes \\ 
		\bottomrule 
	\end{tabular}
\end{center}
\vspace{4pt}
\noindent
When comparing this sample to the sample generated by IncLing in \autoref{sec_problem}, we see that, in contrast, it contains a configuration that covers the fault in the example (\ie{} Configuration~03).
This corresponds to an increase in testing effectiveness.
Compared to the sample produced by ICPL in \autoref{sec_problem}, we can see that it contains only the two real interactions of $P$ and $G$, and thus requires only five configurations, which is an increase in testing efficiency.

	\newcommand{\lesservalue}{\textbf{--}}
\newcommand{\greatervalue}{\textbf{+}}
\newcommand{\samevalue}{\textbf{=}}

\section{Evaluation}

With \pccoverage{} we aim to generate samples for a novel coverage criterion, which we expect to increase the chance of detecting faults in product-based testing.
We are interested in the degree of testing effectiveness and testing efficiency of the \pccoverage{} criterion and our algorithm for \pcsampling{}.
Therefore, we evaluate whether samples generated with \pcsampling{} can detect more faults than samples generated with t-wise interaction sampling.
We also evaluate what degree of \magic{} can be achieved by existing algorithms for t-wise interaction sampling.
Further, we evaluate the sample size (\ie{} testing efficiency) and sampling time (\ie{} sampling efficiency) of our tool \prototype{} (cf. \autoref{sec_concept}).
In summary, we aim to answer the following research questions:
\newlist{rqitemize}{itemize}{3}
\setlist[rqitemize,1]{label=\textbullet,leftmargin=12mm}
\begin{rqitemize}
\item[$RQ_{1}$] Is t-wise presence condition coverage more effective in detecting faults than t-wise interaction coverage for the same value of $t$?
\item[$RQ_{2}$] What degree of presence condition coverage can be achieved using traditional sampling algorithms?
\item[$RQ_{3}$] Is the \emph{testing} of samples more efficient with \pcsampling{} than with $t$-wise interaction sampling?
\item[$RQ_{4}$] Is the \emph{generation} of samples more efficient with \pcsampling{} than with $t$-wise interaction sampling?
\end{rqitemize}

Within our experiments, we compute several samples for different systems using our tool \prototype{} and a selection of different state-of-the-art $t$-wise interaction sampling algorithms and compare the samples with respect to our evaluation criteria.
In the following, we describe the setup for our experiments and our evaluation results.
First, we introduce the algorithms that we compare against each other.
Second, we present the subject systems, for which we generate samples.
Third, we describe our measuring methods for our four evaluation criteria, fault detection, coverage, sample size, and sampling time.
Fourth, we analyze and discuss our results.
Finally, we discuss potential threats to the validity of our evaluation.

\subsection{Algorithms}\label{sec_algos}

We use several state-of-the-art algorithms for $t$-wise interaction sampling as comparison for testing efficiency and effectiveness, which were also used in previous evaluations~\cite{AKT+:GPCE16,ATL+:SoSyM16,MHF:SPLC12}.
First, we employ Chv\'atal~\cite{C:MOR79}, ICPL~\cite{JHF:MODELS11,MHF:SPLC12}, and IncLing~\cite{AKT+:GPCE16} as pure $t$-wise interaction sampling algorithms.
Second, we use \prototype{} as a pure $t$-wise interaction sampling algorithm (cf. \autoref{sec_ppp}), which only uses the feature model as input (\prototype{}-FM).
All of these algorithms compute complete $t$-wise samples for certain values of $t$ using different methods.
Third, we use a random sampling algorithm~\cite{AMK+:GPCE16}.
Instead of aiming for a certain coverage criteria it generates a fixed number of valid random configurations.
Fourth, we include the algorithm PLEDGE~\cite{HPP+:TSE14}, which does not try to achieve a certain $t$-wise interaction coverage, but is based on an evolutionary algorithm to optimize a sample of fixed size such that its contained configurations are as dissimilar as possible.
By increasing dissimilarity, the sample's $t$-wise interaction coverage should also increase.
Although this approach does not guarantee a complete $t$-wise interaction coverage, it aims to increase sampling and testing efficiency while maintaining a reasonably good testing effectiveness.
Finally, we use \prototype{} to compute samples based on presence conditions (\prototype{}-PC).

\subsubsection{Implementation Details}\label{sec_implementation}

The implementation of these algorithm is provided by multiple open-source Java libraries, which we employ in our evalution.
Chv\'atal and ICPL are implemented in the SPLCATool~\cite{MHF:SPLC12}.
IncLing and Random are implemented in FeatureIDE~\cite{AMK+:GPCE16,MTS+17}.
PLEDGE is implemented in a library of the same name~\cite{HPP+:TSE14}.
For all other sampling algorithms we use \prototype, for which we employ our own implementation\footnote{\url{https://github.com/skrieter/evaluation-pc-sampling}}.

We implemented our prototype \prototype{} for \pcsampling{}.
It includes an algorithm for extracting presence conditions from systems that use the \emph{C preprocessor} and the \emph{kbuild} build tool.
\prototype{} is written in Java and employs several other Java libraries to implement its functionality.

For parsing C files and identifying preprocessor statements, we use the tool PCLocator~\cite{KuiterKKLLS18}, which combines several C parsers, such as SuperC, TypeChef, and FeatureCoPP to achieve more accurate results.
This tool computes a presence condition for each line in a source file.
To this end, we analyze every C file (\ie{} files with the file extensions \texttt{.c}, \texttt{.h}, \texttt{.cxx}, and \texttt{.hxx}) in the source directories of the target \ac{spl}.
For this, we exclude special directories that do not contribute to the actual implementation of the system, but contain examples, configuration logic, or header files of system libraries.
As result, we get a list containing propositional formulas for each code block within a C project.
We use this list as input for the \pcsampling{}.
Currently, we did not implement an extraction algorithm for any other variability mechanism.
Thus, in our evaluation, we focus on C projects that use the C preprocessor and kbuild as build system to enable variability.
During the extraction process, we warn the user, if we find presence conditions that contain features that are not on the feature model and vice versa.
For our evaluation, we only consider features that we can find in both, the feature model and the source code.

Within our sampling algorithm, we use the satisfiability solver \emph{Sat4J}~\cite{le2010sat4j} to check for validity of configurations and presence conditions.
Furthermore, we use \emph{KClause}~\cite{oh2019uniform} to extract a feature model for C projects that use Kconfig as configuration tool.

Regarding the random sampling, we use the default random sampling algorithm of FeatureIDE~\cite{AMK+:GPCE16}.
Their implementation is based on \emph{Sat4J} as well and generates configurations by asking the satisfiability solver for a valid configuration using a randomized feature order.
While this algorithm does not generate uniformly distributed random samples, as it is biased by the internal structure of the solver, it is an efficient way to generate a high number of valid configurations.
Note that it is possible for this algorithm to generate a sample that contains duplicate configurations.

\subsubsection{Parameter Details}

As we employ a variety of sampling algorithms in our evaluation, the required parameters differ for most of them.
The only common parameter for every algorithm is the feature model, which specifies the feature dependencies.
Naturally, all algorithms always use the same feature model as input.

Regarding the parameter $t$, not all algorithms support the same values.
IncLing is designed as a strict pair-wise interaction coverage algorithm, and thus only works for $t = 2$.
ICPL supports values for $t$ up to $3$ and Chv\'atal up to $4$.
As described in \autoref{sec_concept}, we can run our algorithm for \pcsampling{} with any value for $t$.
However, \prototype{} currently has a technical limitation that allows to process only up to $2^{31}$ interactions.
To enable a fair comparison, we set the value of $t$ to $t = 2$ for all algorithms.

As Random and PLEDGE do not try to achieve a certain $t$-wise coverage, but just generate a set of valid configurations, it is not possible to set a value for $t$.
Instead, they require to set the size of the sample in advance.
In order to ensure a fair comparison, for PLEDGE, we set the sample size equal to the size of the largest sample computed by any variant of \prototype{} (\ie{} either \prototype{}-FM, \prototype{}-PC, or \prototype{}-Concrete, which ever returned the largest sample).
For Random, we set several sample sizes for each system, ranging from the smallest to the largest sample size produced for every system by any algorithm.

PLEDGE also requires to set a time limit for the evolutionary algorithm.
We decided to compare two different limits, the maximum and minimum time that \prototype{} needs to compute a sample for a particular model (independent from its parameters settings).

For \prototype{} we also have to specify additional parameters beside $t$.
We are able to specify which expressions should be considered for interaction (cf. \autoref{sec_ppp}).
Thus, we test the following setting:
\emph{FM} considers all $t$-wise interactions within a feature model, and thus behaves like other pure $t$-wise interaction sampling algorithms.
\emph{PC} considers all $t$-wise interactions between all presence conditions of a system.
Finally, \emph{Concrete} considers $t$-wise interactions between features, but only includes features that appear in at least one presence condition (\ie{} concrete features).

\subsubsection{Summary}
We compare results from the following algorithms:
\begin{enumerate}
	\item Chv\'atal~\cite{C:MOR79}
	\item ICPL~\cite{JHF:MODELS11,MHF:SPLC12}
	\item IncLing~\cite{AKT+:GPCE16}
	\item PLEDGE-Min (Minimum run time)~\cite{HPP+:TSE14}
	\item PLEDGE-Max (Maximum run time)~\cite{HPP+:TSE14}
	\item \prototype{}-FM (All features within a model)
	\item \prototype{}-PC (All presence conditions)
	\item \prototype{}-Concrete (All concrete features)
	\item Random~\cite{AMK+:GPCE16}
\end{enumerate}

\subsection{Subject Systems}
\begin{table}[t]
	\caption{Subject systems --- features (\textit{\#F}), concrete features (\textit{\#CF}), dependencies (\textit{\#D}), presence conditions (\textit{\#PC}), and the number of clauses (\textit{\#C}) and literals (\textit{\#L}) over all presence conditions.}
	\label{tab_systems}
	\centering
	\begin{tabular}{l|rrr|rrr}
		\toprule
		\textbf{System (Version)} & \multicolumn{3}{c|}{\textbf{Feature Model}} & \multicolumn{3}{c}{\textbf{Presence Conditions}} \\
		& \textit{\#F} & \textit{\#CF} & \textit{\#D} & \textit{\#PC} & \textit{\#C} & \textit{\#L} \\
		\midrule
		fiasco (latest) & 71 & 7 & 120 & 9 & 12 & 14 \\
		axtls (latest) & 95 & 32 & 190 & 90 & 126 & 162 \\
		uclibc-ng (latest) & 270 & 104 & 1,561 & 225 & 315 & 406 \\
		toybox (latest) & 323 & 8 & 90 & 14 & 14 & 14 \\
		BusyBox (1.29.2) & 1,018 & 507 & 997 & 1,020 & 1,475 & 1,975 \\
		Linux (2.6.28.6) & 6,888 & 1,696 & 80,715 & 3,512 & 5,494 & 8,767 \\
		\midrule
		busybox  (1.23.1) & -- & -- & -- & 3,278 & 5,046 & 7,281
\\
		bison (2.0) & -- & -- & -- &   695 & 1,161 &  1,871
\\
		cvs (1.11.17) & -- & -- & -- & 1,495 & 2,491 &  3,785
\\
		libssh (0.5.3) & -- & -- & -- &    393 &  663 &   962
\\
		dia (0.97.2) & -- & -- & -- &    606 &  708 &   810
\\
		libxml2 (2.9.0) & -- & -- & -- &   2,420 & 4,423 &  6,757
\\
		xterm (224) & -- & -- & -- &    796 & 1,302 &  1,859
\\
		lighttpd (1.4.30) & -- & -- & -- &    567 &  875 &  1,219
\\
		libpng (1.5.14) & -- & -- & -- &   1,752 & 3,937 &  7,421
\\
		fvwm (2.4.15) & -- & -- & -- &    777 & 1,482 &  4,075
\\
		irssi (0.8.15) & -- & -- & -- &    318 &  369 &   428
\\
		gnome-keyring (3.14.0) & -- & -- & -- &    453 &  539 &   631
\\
		vim (6.0) & -- & -- & -- &   3,888 & 8,714 & 16,613
\\
		xfig (3.2.4) & -- & -- & -- &    378 &  802 &  1,969
\\
		totem (2.17.5) & -- & -- & -- &    223 &  278 &   332
\\
		gnome-vfs (2.0.4) & -- & -- & -- &    253 &  313 &   373
\\
		cherokee (1.2.101) & -- & -- & -- &   1,128 & 1,589 &  2,077
\\
		bash (4.2) & -- & -- & -- &   3,659 & 6,577 & 10,262
\\
		lua (5.2.1) & -- & -- & -- &    324 &  496 &   714
\\
		gnuplot (4.6.1) & -- & -- & -- &   1,546 & 2,720 &  4,145
\\
		apache (2.4.3) & -- & -- & -- &   1,814 & 2,915 &  4,360
\\
		\bottomrule
	\end{tabular}
\end{table}

Currently, \prototype{} can extract presence conditions from C preprocessor statements.
Thus, we selected real-world open-source systems that use the C preprocessor as a variability mechanism.
In particular, we reused 21 systems from the study of \textcite{MKR+:ICSE16}, which also compared different sampling algorithms in terms of testing effectiveness.
However, most of these systems do not have a separate feature model, which prevents us from taking their feature dependencies into account.
For this reason, we include six real-world open-source systems that use the C preprocessor and the Kconfig tool, namely, \texttt{fiasco} (latest), \texttt{axtls} (latest), \texttt{uclibc-ng} (latest), \texttt{toybox} (latest), \texttt{BusyBox} (version 1.29.2), and \texttt{Linux} (version 2.6.28).
For \texttt{Linux}, we use a feature model for version 2.6.28 provided by She et al.~\cite{SLBWC:ICSE11}.
For all other systems, we extracted the feature models from their Kconfig files using the tool \emph{KClause}~\cite{oh2019uniform}.

In \autoref{tab_systems}, we provide an overview of the all systems.
At the top we show the 6 systems for which we have a feature model and at the bottom the 21 systems from the study of \textcite{MKR+:ICSE16}.
For each respective feature model we show its number of features (\textit{\#F}), concrete features (\textit{\#CF}) (\ie{} features that appear in at least one presence condition), and dependencies (\textit{\#D}).
Regarding the extracted presence conditions, we show the total number of conditions (\textit{\#PC}),  and the number of literals (\textit{\#L}) and clauses (\textit{\#C}) over all presence condition.

\subsection{Evaluation Setup}

\subsubsection{Measuring Fault Detection}

To answer our first research question, we reuse some artifacts from the study of~\textcite{MKR+:ICSE16}.
In the study the authors report known faults in multiple systems and their respective presence conditions\footnote{\url{http://www.dsc.ufcg.edu.br/~spg/sampling/}}.
In this case, if the presence condition of a fault is active under a given configuration, it means that the fault will be present in the corresponding product.
In total, the study presents a list of 75 unique presence conditions.
However, 23 of these conditions contained features that do not occur in the actual source code.
This can be due to abstract features, features that are only used during the build process (\eg{} in Makefiles), or due to features that have a different name in the configuration tool than in the source code.
Therefore, we only used the remaining 52 presence conditions in our evaluation.
Most of these presence conditions represent interaction of degree one, which means that the selection or deselection of a single feature is enough to make them active.
However, the list also contains presence conditions that represent interactions of degree two, three, four, and five.
We show the distribution of presence conditions in \autoref{tab_artifact_pcs} together with an example for each degree of interaction.
Note that, five of the 52 presence conditions also have multiple clauses in their DNF.
For such a case, we consider the number of literals in the smallest clause as degree of interaction for that presence condition.
For instance, the presence condition \texttt{(SHUTDOWN\_SERVER} \texttt{\&\&} \texttt{NO\_SOCKET\_TO\_FD} \texttt{\&\&} \texttt{START\_RSH\_WITH\_POPEN\_RW)} \texttt{||} \texttt{(NO\_SOCKET\_TO\_FD} \texttt{\&\&} \texttt{!SHUTDOWN\_SERVER} \texttt{\&\&} \texttt{START\_RSH\_WITH\_POPEN\_RW)} also represent an interaction of degree three, because it can be activated by the (de)selection of three features.
Thus a complete three-wise interaction coverage would be guaranteed to find this fault.
All in all, the study includes a wide variety of interaction faults with varying degrees of complexity.
\looseness=-1

\begin{table}[t]
	\setlength\tabcolsep{5pt}
	\caption{Overview of presence conditions of faults used from \textcite{MKR+:ICSE16}.}
	\label{tab_artifact_pcs}
	\centering
	\begin{tabular}{rrl}
		\toprule
		\textbf{Degree} & \textbf{Count}  & \textbf{Example}  \\
		\midrule
		1 & 34 & \texttt{!ENABLE\_FEATURE\_SYSLOG} \\[2pt]
		2 & 11 & \texttt{ENABLE\_FEATURE\_EDITING \&\& !ENABLE\_HUSH\_INTERACTIVE} \\[2pt]
		3 & 4 & \texttt{ENABLE\_FEATURE\_GETOPT\_LONG \&\& !ENABLE\_FEATURE\_SEAMLESS\_LZMA} \\
		 &  & ~~~\texttt{\&\& !ENABLE\_FEATURE\_TAR\_LONG\_OPTIONS} \\[2pt]
		4 & 2 & \texttt{!FEAT\_GUI\_W32 \&\& !PROTO \&\& !FEAT\_GUI\_MOTIF \&\& !FEAT\_GUI\_GTK} \\[2pt]
		5 & 1 & \texttt{!FEAT\_GUI\_W32 \&\& !FEAT\_GUI\_GTK \&\& !FEAT\_GUI\_MOTIF} \\
		 & & ~~~\texttt{\&\& !FEAT\_GUI\_ATHENA \&\& !FEAT\_GUI\_MAC \&\& FEAT\_GUI} \\
		\midrule
		$\sum$ & 52 & \\
		\bottomrule
	\end{tabular}
\end{table}

We use the list of presence conditions to check whether samples generated for theses systems do cover each fault in at least one configuration.
To this end, we generate samples for each of these systems with \prototype{}-PC (\ie{} presence condition coverage) and \prototype{}-Concrete (\ie{} interaction coverage) for $t=1$ and $t=2$.
We then count how many reported faults are covered by each sample.
To determine whether a fault is covered, we check if there exists at least one configuration in the sample that satisfies the corresponding presence condition of the fault.

Both algorithms are susceptible to the order of features or order of presence conditions that are provided as input, meaning that they will produce different results for different feature orders.
Thus, we evaluate both algorithms using multiple iterations with a randomized feature order.
In detail, we execute all algorithms 100 times, each time shuffling the feature order.
To enable a fair comparison we use the same 100 randomized feature orders for each algorithm.
A number of 100 iterations is an empirical value for our evaluation that provides a good trade-off between effort and accuracy.

Note, that we do not use any of the other algorithms in this experiment, as we do not have feature models for these systems.
The lack of a feature model for a system also means that the configurations within a sample may be invalid according to the feature dependency of the system.
However, without a feature model we are not able to test this.
 
\subsubsection{Measuring Coverage}

We compute the coverage achieved by every sample with regard to two different coverage criteria, pair-wise interaction coverage (\emph{FM}) and pair-wise presence condition coverage (\emph{PC}).
We consider a sample and, consequently, its sampling algorithm to be more effective the higher its coverage, as it potentially exposes more faults in the code.

Similar to the previous experiment, all used sampling algorithms are susceptible to the feature order in a feature model.
Thus, again, we execute all algorithms 100 times, each time shuffling the feature order.
In addition, we execute Random 10 times for each feature order, which results in 1,000 iterations for each system.
For \texttt{Linux}, we only use 5 iterations of the experiment, as most algorithms take several hours to compute just one sample.

\subsubsection{Measuring Sample Size}

Regarding testing efficiency, we count the number of configurations in each sample computed by each algorithm.
We do not consider the time required to run any actual test cases of a particular system.
We do not consider the time required to run any test cases of a particular system, as this time is depended on the actual test cases for each product and the general testing approach.
Nevertheless, we can assume that the testing time increases with the number of configuration in a sample, and thus, in general, a smaller sample will lead to a smaller testing time.
Analogous to measuring coverage, we execute each algorithm, except Random, 100 times and randomize the feature order.
Random is again run 1,000 times for each sample size.

\subsubsection{Measuring Sampling Time}

For measuring sampling efficiency, we take the time that is needed for generating a sample with each algorithm.
Each experiment runs on an own JVM, in order to mitigate any side effects (\eg{} just-in-time compilation).
As our algorithm requires additional information from the source code (\ie{} the presence conditions), we differentiate between the time needed to extract the presence conditions from the source code and the time to actually generate the sample.
This is relevant, as the extraction process only needs to be run once for each system.
Though it takes some time to analyze the source code, the resulting presence conditions for each file can be saved for later reuse.
For instance, if we compute samples for different values of $t$, we only need to run the extraction process once.

\subsubsection{Computing Environment}

We run all algorithms on the same computing environment, with the following specifications:
\textit{CPU}: Intel Core i5-8350U, \textit{Memory}: 16\,GB, \textit{OS}: Manjaro (Arch Linux), \textit{Java}: OpenJDK 15.0.2, \textit{JVM Memory}: Xmx: 14\,GB, Xms: 2\,GB.

\subsection{Evaluation Results}

For brevity, we primarily present figures showing aggregated data over our measurement results.
All data and a tabular overview can be found online.\footnote{\url{https://github.com/skrieter/evaluation-pc-sampling/tree/master/results}}
We structure our findings according to our four research questions, that is fault detection, coverage, testing efficiency, and sampling efficiency.
Afterwards, we analyze and discuss our results.

\subsubsection{Faults Covered}

\begin{table}[t]
	\setlength\tabcolsep{5pt}
	\caption{Faults covered across all 21 systems from \textcite{MKR+:ICSE16}, including aggregated sample size and sampling time over all systems.}
	\label{tab_faults}
	\centering
	\begin{tabular}{lr|rrr|rrr|rr}
		\toprule
		\textbf{Algorithm} & \textbf{t} & \multicolumn{3}{c|}{\textbf{Size}} & \multicolumn{3}{c|}{\textbf{Time (s)}} & \multicolumn{2}{c}{\textbf{Faults Covered}} \\
		& & $\varnothing$ & \emph{Min} & \emph{Max} & $\varnothing$ & \emph{Min} & \emph{Max} & \emph{Yes} & \emph{No} \\
		\midrule
		PRESICE-PC & 1 & 7.5 & 4 & 14 & 0.3 & 0.2 & 0.6 & 41 & 11 \\
		PRESICE-Concrete & 1 & 2.0 & 2 & 2 & 0.3 & 0.2 & 0.4 & 36 & 16 \\
		PRESICE-PC & 2 & 65.7 & 22 & 167 & 5.6 & 0.3 & 38.8 & 51 & 1 \\
		PRESICE-Concrete & 2 & 16.7 & 12 & 24 & 0.9 & 0.2 & 3.7 & 47 & 5 \\
		\bottomrule
	\end{tabular}
\end{table}

We present the results of our first experiment in \autoref{tab_faults}.
For each algorithm and value for $t$, we show the number of faults that are covered or not covered by the produced samples across all systems.
The number of covered faults is the \emph{minimum} number over all 100 iterations, meaning that if any of the 100 samples for a system was not able to cover a particular fault it is \emph{not} counted as covered.
Analogous, the number of not covered faults is the \emph{maximum} number over all 100 iterations.
In addition, we report the aggregated sample size and sampling time over all systems.
For both values, we report its minimum, maximum and average over all 21 systems and 100 iterations.

Of the 52 faults, which we investigated, we see that for both values of $t$ \prototype{}-PC is able to detect more faults than \prototype{}-Concrete (\ie{} 31 vs. 36 for $t=1$ and 51 vs. 47 for $t=2$).
The presence condition that belongs to the fault that could not always be covered by \prototype{}-PC with $t=2$ is the following:\\
\hspace*{20pt}
\texttt{ENABLE\_HUSH\_CASE} \texttt{\&\&} \texttt{ENABLE\_FEATURE\_EDITING\_SAVE\_ON\_EXIT}\\
\hspace*{30pt}
\texttt{\&\&} \texttt{ENABLE\_HUSH\_INTERACTIVE} \texttt{\&\&} \texttt{!ENABLE\_FEATURE\_EDITING}\\
This presence condition is of degree four, and thus, when using t-wise interaction sampling, is only guaranteed to be found with $t\geq4$.

On the other hand, we can also see that on average \prototype{}-PC produced larger samples than \prototype{}-Concrete (\ie{} 7.5 vs. 2.0 for $t=1$ and 65.7 vs. 16.7 for $t=2$).
Thus, the higher fault detection may also be a result of the larger sample sizes.
However, as we pointed out before, we do not use a feature model for this experiment.
Therefore, there are no restrictions on the configuration space, which can lead to a lower sample size.
Furthermore, as we see in later experiments, a feature model, which may also include abstract features, leads to a larger sample size than considering only concrete features.

\subsubsection{Achieved Coverage}

\begin{table}
	\caption{Relative mean sample size, mean sampling time, and mean coverage aggregated over all 6 systems with a feature model.}
	\label{tab_coverage}
	\centering
	\begin{tabular}{l|rrrrr}
		\toprule
		\textbf{Algorithm} & $\varnothing$\textbf{Time (\%)} & $\varnothing$\textbf{Size (\%)} & \multicolumn{2}{c}{$\varnothing$\textbf{Coverage (\%)}} \\
		&  &  & \emph{FM} & \emph{PC} \\
		\midrule
		PRESICE-FM &  100.0 & 100.0 & 100.0 &  98.6 \\
		PRESICE-PC &   59.3 &  73.9 &  79.1 & 100.0 \\
		PRESICE-Concrete &   36.1 &  16.6 &  61.7 &  62.9 \\
		ICPL &  319.5 & 132.7 & 100.0 &  97.9 \\
		Chv\'atal & 1,046.7 & 131.3 & 100.0 &  98.0 \\
		IncLing &   53.6 & 153.7 & 100.0 &  99.3 \\
		PLEDGE-Min &   51.5 & 122.0 &  98.8 &  97.1 \\
		PLEDGE-Max &  118.2 & 122.0 &  98.8 &  97.1 \\
		\bottomrule
	\end{tabular}
\end{table}

\begin{figure}
\includegraphics[width=.95\linewidth]{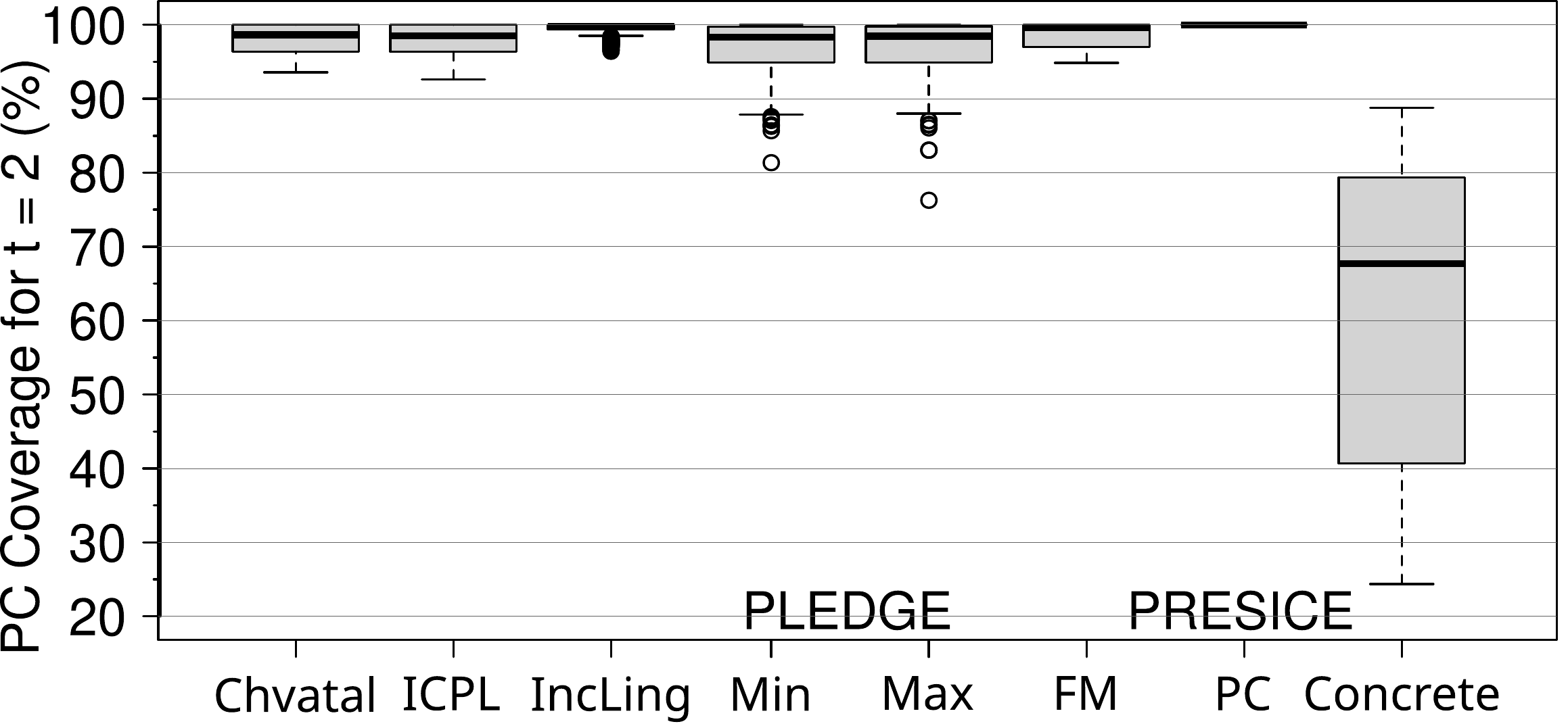}
\caption{Pair-wise presence condition coverage aggregated over all systems.}
\label{fig_effectiveness}
\end{figure}

In \autoref{tab_coverage}, we show a comparison of the coverage for different criteria for all algorithms.
These values are aggregated over all systems and all experiments using the arithmetic mean.
We show a more detailed plot of the coverage criterion \emph{PC} over all systems and experiments in \autoref{fig_effectiveness} using boxplots.
In addition, we performed paired t-tests to test whether the difference in achieved coverage by the different algorithms is significant.
We present the results of the statistical tests in \autoref{tab_stat_coverage}.
In this table, we compare the coverage of all three variants of PRESICE with the coverage of all other algorithms for the two coverage criteria \emph{FM} and \emph{PC} with $t = 2$.
The symbol \samevalue{} indicates that there is no significant difference (\ie{} $p > 0.05$) in the achieved coverage between both algorithms.
The symbol \lesservalue{} indicates that the coverage achieved by the variant of PRESICE is significantly lower than the coverage of the other algorithm (\ie{} $p < 0.001$).
Analogous, the symbol \greatervalue{} indicates the coverage of PRESICE is significantly higher (\ie{} $p < 0.001$).

Regarding the coverage criterion \emph{FM}, we can see that only the $t$-wise interaction sampling algorithms (Chv\'atal, ICPL, IncLing, PRESICE-FM) are able to achieve a 100\% coverage.
Both, PRESICE-PC and PRESICE-Concrete achieve a significant lower \emph{PC} coverage.
On the other hand, only PRESICE-PC is able to achieve a 100\% coverage criterion \emph{PC}.
All other algorithms produce samples with a significant lower \emph{PC} coverage on average.
Still, many algorithms (Chv\'atal, ICPL, IncLing, PLEDGE, PRESICE-FM) achieve a rather high average \emph{PC} coverage of over 97\%.

\begin{table}
	\caption{Results of the paired t-tests for difference in \emph{FM} and PC coverage between PRESICE and other algorithms.}
	\label{tab_stat_coverage}
	\centering
	\begin{tabular}{l|ccccc}
		\toprule
		\textbf{Algorithm} & \textbf{ICPL} & \textbf{Chvatal} & \textbf{IncLing} & \multicolumn{2}{c}{\textbf{Pledge}} \\
		&  &  &  & \textit{Min} & \textit{Max}  \\
		\midrule
\textit{\textbf{FM-Coverage}} &&&&&\\
~~\textbf{PRESICE-FM} & \samevalue{} & \samevalue{} & \samevalue{} & \greatervalue{} & \greatervalue{}  \\
~~\textbf{PRESICE-PC} & \lesservalue{} & \lesservalue{} & \lesservalue{} & \lesservalue{} & \lesservalue{}  \\
~~\textbf{PRESICE-Concrete} & \lesservalue{} & \lesservalue{} & \lesservalue{} & \lesservalue{} & \lesservalue{}  \\
\midrule
\textit{\textbf{PC-Coverage}} &&&&&\\
~~\textbf{PRESICE-FM} & \greatervalue{} & \greatervalue{} & \lesservalue{} & \greatervalue{} & \greatervalue{}  \\
~~\textbf{PRESICE-PC} & \greatervalue{} & \greatervalue{} & \greatervalue{} & \greatervalue{} & \greatervalue{} \\
~~\textbf{PRESICE-Concrete} & \lesservalue{} & \lesservalue{} & \lesservalue{} & \lesservalue{} & \lesservalue{}  \\
		\bottomrule
\end{tabular}
\end{table}

\subsubsection{Sample Size}

\begin{figure}[t]
\includegraphics[width=.95\linewidth]{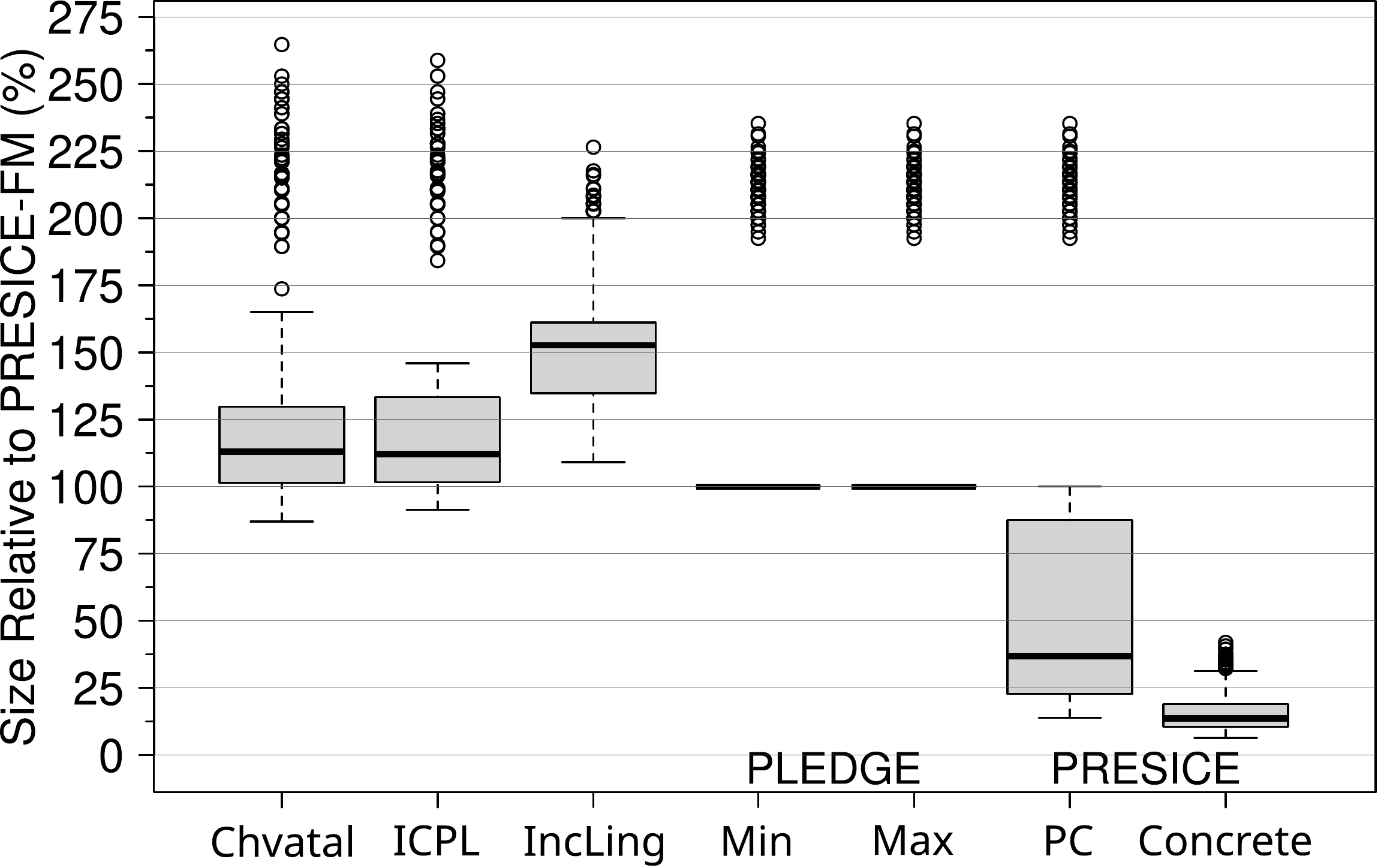}
\caption{Sample size relative to \prototype{}-FM for all 6 systems with a feature model.}
\label{fig_size_per_t}
\end{figure}

In \autoref{tab_coverage}, we show a comparison of the sample sizes for all algorithms.
Again, the values are aggregated over all systems and all experiments using the arithmetic mean.
As the actual sample size is dependent on the subject system, we normalized the sample size for every experiment using the sample size of \prototype{}-FM as 100\%.
In \autoref{fig_size_per_t}, we depict the sample size for all algorithms in more detail using boxplots.
Additionally, we show the absolute values of the mean sample size per system for \prototype{}-FM and \prototype{}-PC in \autoref{tab_absolute}.
Furthermore, we performed paired t-tests to test whether the difference in sample size computed by the different algorithms is significant and show the results in \autoref{tab_stat_size}.
In this table, we compare the sample size of all three variants of \prototype{} with the sample size of all other algorithms.
The symbol \lesservalue{} indicates that the sample size computed by the variant of \prototype{} is significantly smaller than the sample size of the other algorithm (\ie{} $p < 0.001$).
Analogous, the symbol \greatervalue{} indicates the sample size of \prototype{} is significantly greater (\ie{} $p < 0.001$).

We can observe that on average all algorithms that use presence conditions as input produce significantly smaller samples than algorithm that only use the feature model.
An exception is the system \texttt{BusyBox}, for which \prototype{}-PC produces samples that are about two times larger than the sample of \prototype{}-FM.

\begin{table}[t]
	\caption{Results of the paired t-tests for sample size difference between PRESICE and other algorithms.}
	\label{tab_stat_size}
	\centering
	\begin{tabular}{l|ccccc}
		\toprule
		\textbf{Algorithm} & \textbf{ICPL} & \textbf{Chvatal} & \textbf{IncLing} & \multicolumn{2}{c}{\textbf{Pledge}} \\
		&  &  &  & \textit{Min} & \textit{Max} \\
		\midrule
		PRESICE-FM & \lesservalue{} & \lesservalue{} & \lesservalue{} & \lesservalue{} & \lesservalue{} \\
		PRESICE-PC & \lesservalue{} & \lesservalue{} & \lesservalue{} & \lesservalue{} & \lesservalue{}  \\
		PRESICE-Concrete & \lesservalue{} & \lesservalue{} & \lesservalue{} & \lesservalue{} & \lesservalue{}  \\
		\bottomrule
	\end{tabular}
\end{table}

\begin{table}[t]
	\caption{Absolute mean sample size and mean sampling time for PRESICE-FM and PRESICE-PC for all 6 systems with a feature model.}
	\label{tab_absolute}
	\centering
	\begin{tabular}{l|rr|rr|r}
		\toprule
		\textbf{System} & \multicolumn{2}{c|}{$\varnothing$\textbf{Size}} & \multicolumn{2}{c|}{$\varnothing$\textbf{Time (s)}} & $\varnothing$\textbf{Extract (s)} \\
		& \textbf{FM} & \textbf{PC} & \textbf{FM} & \textbf{PC} & \\
		\midrule
		fiasco &  21.5 &   5.4 &  1.0 &    0.6 &   0.7 \\
		axtls &  32.3 &  27.3 &  1.4 &    1.3 &   1.4 \\
		uclibc-ng & 362.4 &  54.5 &    6.8 &    3.3 &  0.8 \\
		toybox &  18.4 &   6.5 &    3.6 &    0.7 &   4.5 \\
		BusyBox &  37.6 &  79.1 &   28.6 &   20.7 &  2.1 \\
		Linux & 493.4 & 189.4 & 8,938.4 & 1,248.6 & 64.5 \\
		\bottomrule
	\end{tabular}
\end{table}

\subsubsection{Correlation Between Coverage and Sample Size}

\begin{figure}
\includegraphics[width=.95\linewidth]{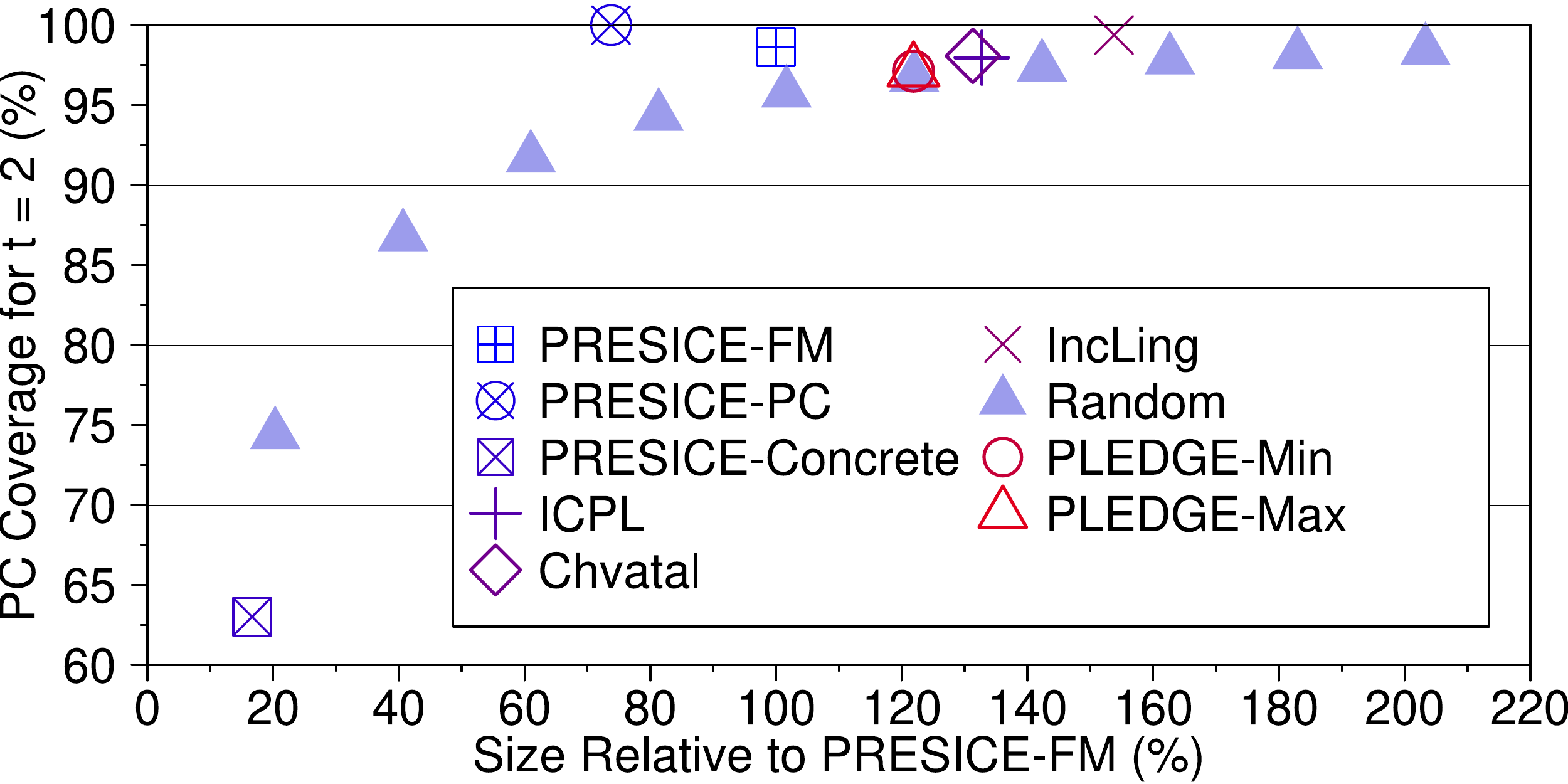}
\caption{Presence condition coverage compared to sample size for all algorithms aggregated over all 6 systems with a feature model.}
\label{fig_effectiveness_size}
\end{figure}

To illustrate the correlation between sample size and testing effectiveness, we show, in \autoref{fig_effectiveness_size}, a comparison of the coverage criterion \emph{PC} with $t = 2$ for all algorithms for different configuration sizes.
On the x-axis, we show the sample size relative to the sample size of \prototype{}-FM (\ie{} being 100\%).
On the y-axis, we show coverage in \% for \emph{PC} with $t = 2$.
Each data point represents the average for all samples per algorithm and system.
Random acts as a base line in this diagram, as it does not aim for a certain coverage criterion.
We can see a clear correlation between sample size and the coverage criterion \emph{PC} (\ie{} increasing the sample size leads to higher coverage on average).
Further, we can see that \prototype{} can reach a 100\% coverage with a substantially smaller sample than all other tested algorithms for most cases.

In addition, we calculate the Spearman's rank correlation coefficient between the degree of coverage and the sample size for all algorithms.
For the coverage criterion \emph{PC}, we get a significant positive correlation of $\approx 0.157$ ($p < 0.001$).
Similarly, for the coverage criterion \emph{FM}, we also get a significant positive correlation of $\approx 0.2$ ($p < 0.001$).

\subsubsection{Sampling Time}

\begin{figure}
\includegraphics[width=.95\linewidth]{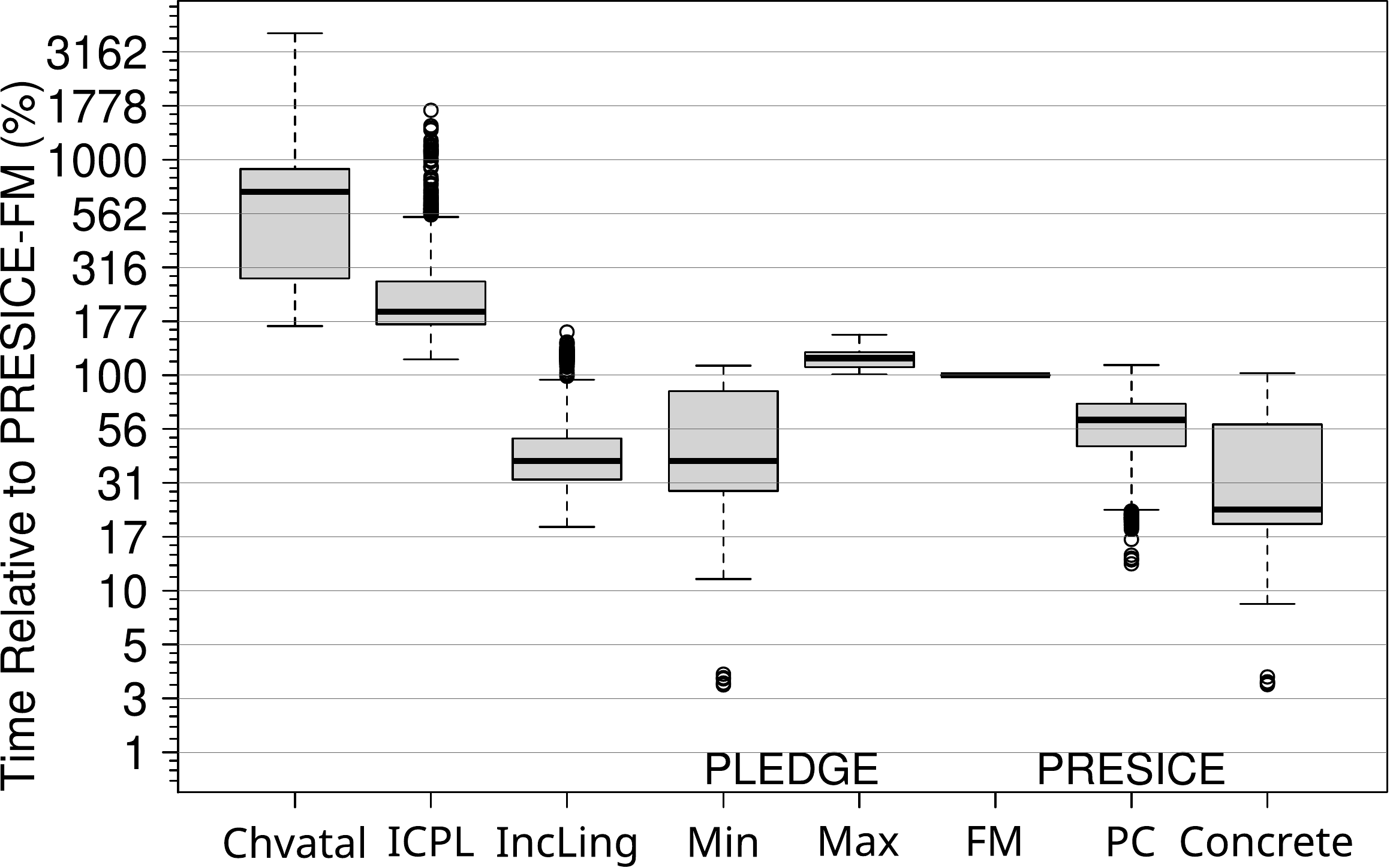}
\caption{Sampling times of all algorithms relative to \textit{\prototype{}-FM} on the same system.}
\label{fig_time}
\end{figure}

In \autoref{tab_coverage}, we show the average sampling time for all algorithms, aggregated over all systems using the arithmetic mean and relative to the sampling time of \prototype{}-FM.
In addition, we depict the sampling time for all algorithms in more detail in \autoref{fig_time} relative to the sampling time of \prototype{}-FM.
Here, we only show the pure sampling time, excluding the time needed for extracting presence conditions.
As we described in \autoref{sec_concept}, the extraction process is dependent on the employed variability mechanism, but only needs to be executed once, if the implementation artifacts do not change.
In \autoref{tab_absolute}, we show for each system the absolute time in seconds required for \prototype{} to extract the presence conditions.
Additionally, in \autoref{tab_absolute}, we show the absolute values for the mean sampling time of \prototype{}-FM and \prototype{}-PC.
We performed paired t-tests to test whether the difference in sampling time required by the different algorithms is significant.
In \autoref{tab_stat_time}, we show the results of these statistical tests by comparing the sampling time of all three variants of \prototype{} with the sample size of all other algorithms.
The symbol \lesservalue{} indicates that the sampling time computed by the variant of \prototype{} is significantly smaller than the sampling time of the other algorithm (\ie{} $p < 0.001$).
Analogous, the symbol \greatervalue{} indicates the sampling time of \prototype{} is significantly greater (\ie{} $p < 0.001$).

We can see that the algorithms that consider presence conditions are significantly faster on average than some algorithms using only the feature model (ICPL, and Chv\'atal, and PLEDGE-Max).
However, IncLing and PLEDGE-Min significantly outperform \prototype{}-FM and are also on average significantly faster than \prototype{}-PC.
Regarding presence condition extraction time, we can see that in most cases it is similar to the sampling time.
In case of Linux and BusyBox the extraction time even is substantially lower than the sampling time.

\begin{table}
	\caption{Results of the paired t-tests for difference in sampling time between PRESICE and other algorithms.}
	\label{tab_stat_time}
	\centering
	\begin{tabular}{l|ccccc}
		\toprule
		\textbf{Algorithm} & \textbf{ICPL} & \textbf{Chvatal} & \textbf{IncLing} & \multicolumn{2}{c}{\textbf{Pledge}} \\
		&  &  &  & \textit{Min} & \textit{Max} \\
		\midrule
		PRESICE-FM & \lesservalue{} & \lesservalue{} & \greatervalue{} & \greatervalue{} & \lesservalue{}  \\
		PRESICE-PC & \lesservalue{} & \lesservalue{} & \greatervalue{} & \greatervalue{} & \lesservalue{}  \\
		PRESICE-Concrete & \lesservalue{} & \lesservalue{} & \lesservalue{} & \lesservalue{} & \lesservalue{} \\
		\bottomrule
	\end{tabular}
\end{table}

\subsection{Discussion}

Regarding $RQ_1$, we found in our experiments that samples with a 100\% \pccoverage{} were able to cover more faults than samples with 100\% $t$-wise interaction coverage.
These results indicate that \pccoverage{} is indeed able to detect more faults than $t$-wise interaction coverage for the same value of $t$.

For $RQ_2$, we can see that only \prototype{}-PC is able to guarantee a 100\% \pccoverage{}.
Nevertheless, most of the other sampling algorithms, such as Chv\'atal, ICPL, IncLing, and PLEDGE achieve a high pair-wise presence conditions coverage (\ie{} over 97\% on average).
While these results indicate an already good \pccoverage{} with traditional sampling algorithms, it also shows that samples from these algorithms most likely miss some interactions between code blocks, which then cannot be tested.
Furthermore, the high coverage of these algorithms might be due to the relatively large sample size compared to \prototype{}-PC.
Notably, random sampling with similar sample sizes also yields a similar coverage as the traditional sampling algorithms.
In summary, it is possible for other algorithms to achieve a high \pccoverage{}, though it cannot be guaranteed.
However, the high \pccoverage{} of the traditional sampling algorithms seems to be caused by their larger sample sizes.

Concerning $RQ_3$, we observe that for all systems, except \texttt{BusyBox}, \prototype{}-PC generates smaller samples, than Chv\'atal, ICPL, IncLing, and \prototype{}-FM.
This may be caused by the relatively low number of concrete features for some systems, which facilitates the coverage of presence conditions within a configuration.
The larger sample size of \texttt{BusyBox} may be caused due to a high number of mutually exclusive presence conditions.
When considering presence conditions, we see that even for systems with more presence conditions than features (\eg{} \texttt{axtls}, \texttt{uclibc-ng}, and \texttt{Linux}) \prototype{}-PC is able to generate smaller samples.
Moreover, the results of \prototype{}-Concrete show that it is crucial to not just consider concrete feature, which yields smaller samples using $t$-wise interaction algorithms, but only reaches a \emph{PC} coverage of about 67\% on average.
In summary, for most cases \pcsampling{} produce relatively small samples, which may increase its testing efficiency.

Regarding $RQ_4$, we see that the sampling time of \prototype{}-PC is relatively small and even outperforms some $t$-wise interaction sampling algorithms.
The additional time for extracting presence conditions is similar to the sampling time for the smaller systems and even neglectable for larger systems such as \texttt{Linux}.
In summary, the initial generation of samples with \prototype{}-PC is only slightly less efficient than with traditional sampling algorithms due to the necessary extraction of presence conditions.

To conclude, with \prototype{} we are able to efficiently generate relatively small samples for \pccoverage{}.
In addition, our results indicate that \pccoverage{} is able to increase the fault-detection rate for product-based testing.
Thus, we may be able to increase the testing efficiency and testing effectiveness by using \pcsampling{}.

\subsection{Threats to Validity}

We are aware that our evaluation might suffer from some biases that may threaten its validity.
First, we are using a rather small set of systems with feature models in our evaluation.
Thus, the results might not scale to other systems with more features or presence conditions.
However, we reused systems from other research and try to get a wide range of feature model sizes.

Second, we do not evaluate the actual testing effectiveness of our approach, but only compare it to other algorithms with respect to our own coverage criterion.
As we proposed the coverage criterion of \pccoverage{} ourself, this may create an unfair comparison with other algorithms that aim for different coverage criteria.
We do evaluate whether samples with 100\% \pccoverage{} are able to detect some known faults in several systems.
However, as this is only a small set of faults, it may hamper the generalizability of our results.
Nevertheless, we tried to include faults with varying degrees of interaction and complex presence conditions to mitigate this bias.

Third, we aim to evaluate the concept of \pcsampling{}.
However, we only use one particular sampling algorithm to cover interactions between presence conditions.
It may be the case that the concept works better or worse when employing other sampling algorithms or different heuristics.
Likewise, there is the chance of implementation bugs that may bias the result.
To this end, we use automated tests to ensure that the samples we receive from \prototype{} are valid and indeed achieve a \pccoverage{} of 100\%.
Furthermore, we used one particular algorithm for extracting presence conditions from C preprocessor \acp{spl}, which has limitations that may affect the correctness of the extracted presence conditions.
However, the tool on which our algorithm is based employs three established tools for analyzing C preprocessor annotations in order to achieve more reliable results.

Finally, as we are using randomized features orders and random sampling, our results may be affected by a random bias.
We tried to mitigate this by using multiple iterations of all experiments.

	\section{Related Work}\label{SEC_Related}

There exist many different approaches to sampling an \ac{spl}~\cite{TAK+:CSUR14,VAT+:SPLC18}.
Like many other sampling algorithms, our sampling algorithm uses a greedy strategy to compute a minimal sample~\cite{AMS+:TOSEM18,AKT+:GPCE16,EB+:ITNG11,HLHE:VaMoS13,JHF:MODELS11,MHF:SPLC12,JHF+:MODELS12,KBK:AOSD11,KBBK:RV10,KSS:VariComp13,LKA+:ESECFSE13,OMR:SPLC10,RBR+:SPLC15,SCD:FASE12,TDS+:ATC14}.
In fact, our sampling algorithm is based on YASA~\cite{KTSSL20}, which is similar to the algorithm IPOG~\cite{LK+:ECBS07}, as it also starts with an empty sample and iteratively adds literals.
However, our sampling algorithm can cope with more general inputs, and thus is able to process arbitrary presence conditions.
Furthermore, there also exist many algorithms that employ meta-heuristic strategies~\cite{CR:IJAST14,FKPV:CEC16,EBG:CAiSE12,FLS+:CIM17,FLV:SBES17,HPL:SBSE14,HPP+:SPLC13,LHF+:CEC14,MGSH:SPLC13,MFV:JSERD16,DPL+:VaMoS15} or other strategies~\cite{PSK+:ICST10} to cope with the variability explosion problem.
These approaches can be seen as complementary to our concept, as we may also adapt meta-heuristic sampling to support presence conditions.

There already exist sampling strategies that consider other inputs in addition to a feature model~\cite{VAT+:SPLC18}.
Analogous to our approach, there are sampling algorithms that consider implementation artifacts~\cite{LKA+:ESECFSE13,SCD:FASE12,TLD:OSR12,TDS+:ATC14} and test artifacts~\cite{KBK:AOSD11,KBBK:RV10} to compute a sample.
While these algorithms also consider the underlying \ac{spl} we are the first to combine presence conditions for implementation artifacts with regular t-wise interaction sampling to achieve higher effectiveness.

	\section{Conclusion \& Future Work}\label{SEC_Conclusion}

With \pccoverage{}, we present a new coverage criterion for \acp{spl} that considers the actual implementation artifacts by looking at their presence conditions.
Further, we describe a $t$-wise sampling algorithm that covers presence conditions instead of features and implement it for systems that use the C preprocessor.
We test the fault-detection rate of \pccoverage{} in comparison to $t$-wise interaction coverage.
We also test our implementation by comparing it to existing sampling algorithms with regard to achieved coverage, sampling size, and sampling time.
We find that \pccoverage{} is able to detect more faults for a given $t$ and that \pcsampling{} produces mostly smaller samples compared to $t$-wise interaction sampling, while guaranteeing a 100\% \pccoverage{}.
Therefore, we suspect that \pcsampling{} has the potential to increase both, testing effectiveness and testing efficiency.

Regarding future work, there are several research aspects, we would like to investigate further.
To begin with, we want to investigate whether the results we achieved in our evaluation for testing effectiveness and efficiencies also scale to large systems with more or more complex presence conditions.
In addition, we also want to evaluate the impact on testing effectiveness and efficiency, when grouping presence conditions as outlined in \autoref{sec_ppp}.
Furthermore, there are many sampling strategies beside $t$-wise interaction sampling, for instance all-yes, all-no, 1-enabled, and 1-disabled.
Each strategy uses a heuristic to trade-off testing efficiency against overall testing effectiveness.
We can easily adapt our coverage criterion and sampling algorithm to consider these strategies for presence condition coverage instead of $t$-wise.
Thus, we want to investigate, whether there is a benefit in using one of these strategies in combination with presence conditions.

	\section*{Acknowledgments}
	We thank Tobias Heß for his valuable feedback.
	
\printbibliography

\end{document}